\documentclass{aa}  

\usepackage{graphicx}
\usepackage{subfig}
\usepackage{bm}
\usepackage{threeparttable}
\usepackage{txfonts}
\bibpunct{(}{)}{;}{a}{}{,}
\begin{document}

   \title{An automated classification approach to ranking photospheric proxies of magnetic energy build-up}
   \titlerunning{Ranking photospheric proxies of magnetic energy build-up}

   \author{A. Al-Ghraibah\inst{1}
          \and
          L. E. Boucheron\inst{1}
          \and
          R. T. J. McAteer\inst{2}
          }

   \institute{Klipsch School of Electrical \& Computer Engineering, New Mexico State University, Las Cruces, NM 88003, USA\\
              \email{amani@nmsu.edu, lboucher@nmsu.edu}
         \and
             Department of Astronomy, New Mexico State University, Las Cruces, NM 88003, USA\\
             \email{mcateer@nmsu.edu}
             }

   \date{Received ; accepted }

 
  \abstract
   {}
   {We study the photospheric magnetic field of $\sim$2000 active regions over solar cycle 23 to search for parameters that may be indicative of energy build-up and its subsequent release as a solar flare in the corona.}
   {We extract three sets of parameters: (1) snapshots in space and time: total flux, magnetic gradients, and neutral lines; (2) evolution in time: flux evolution; and (3) structures at multiple size scales: wavelet analysis. This work combines standard pattern recognition and classification techniques via a relevance vector machine to determine (i.e., classify) whether a region is expected to flare ($\ge$C1.0 according to GOES).  We consider classification performance using all 38 extracted features and several feature subsets.  Classification performance is quantified using both the true positive rate (the proportion of flares correctly predicted) and the true negative rate (the proportion of non-flares correctly classified).  Additionally, we compute the true skill score which provides an equal weighting to true positive rate and true negative rate and the Heidke skill score to allow comparison to other flare forecasting work.}
   {We obtain a true skill score of $\sim$0.5 for any predictive time window in the range 2 to 24~hours, with a true positive rate of $\sim0.8$ and a true negative rate of $\sim0.7$. These values do not appear to depend on the predictive time window, although the Heidke skill score ($<$0.5) does. Features relating to snapshots of the distribution of magnetic gradients show the best predictive ability over all predictive time windows. Other gradient-related features and the instantaneous power at various wavelet scales also feature in the top five (of 38) ranked features in predictive power. It has always been clear that while the photospheric magnetic field governs the coronal non-potentiality (and hence likelihood of producing a solar flare), photospheric magnetic field information alone is not sufficient to determine this in a unique manner. Furthermore we are only measuring proxies of the magnetic energy build up. We are still lacking observational details on why energy is released at any particular point in time. We may have discovered the natural limit of the accuracy of flare predictions from these large scale studies.}
   {}


   \maketitle
%

\section{Introduction}
\label{intro}

Solar flares are the result of magnetic energy release in the corona. As such, we require coronal magnetic field measurements in order to fully understand how this energy is built up (over days) and released (over minutes to hours). However, as such measurements are currently unavailable, much research has focused on inferring the coronal magnetic field structure from those data that are available, namely the photospheric magnetic field. It is assumed that turbulent motions on the surface of the Sun (the photosphere) twist and wind up the magnetic field in the corona.  A complex photosphere causes a complex corona, a complex corona stores energy, and this stored energy is released as solar flares~\citep{mcateer2010}. There have been many searches for a connection between coronal activity (flares) and the photospheric magnetic field~\citep{ahmed2013,falconer2011,mason2010,yuan2010,huang2010,jing2010,yu2010a,yu2010b,welsch2009,ireland2008,conlon2008,georgoulis2007,leka2007,schrijver2007,wang2006,barnes2006,guo2006,jing2006,mcateer2005,abramenko2005,meunier2004,abramenko2003,leka2003b,hagyard1999,zhang1994}. Despite such research, a fully consistent and causal relationship has not yet been found (e.g., see conclusions in \citet{mason2010}, \citet{leka2007}, and \citet{hagyard1999}).  Nevertheless, there is optimism that photospheric measures may yield some insight into imminent eruptive behavior (e.g., \citet{falconer2011,yu2010a,schrijver2007,guo2006,jing2006,abramenko2005,leka2003b}). 

Many of the studies of the photospheric magnetic field extract a single parameter, or a few ($\le 7$) parameters, and look for a relation to solar flare activity~\citep{falconer2011,mason2010,yuan2010,huang2010,jing2010,yu2010a,ireland2008,conlon2008,georgoulis2007,schrijver2007,wang2006,guo2006,jing2006,mcateer2005,abramenko2005,meunier2004,abramenko2003,hagyard1999,zhang1994}. By using full vector magnetograms~\citep{jing2010,leka2007,barnes2006,leka2003b,hagyard1999,zhang1994} to probe the transverse field it is possible to extract a larger number of parameters of magnetic complexity, albeit at the consequence of smaller datasets. Many of these studies focus on a predictive time window of 24 hours~\citep{ahmed2013,falconer2011,yuan2010,jing2010,leka2007,schrijver2007,mcateer2005} although it is not clear that any of the extracted parameters are optimum for a 24-hour predictive time window. To the best of our knowledge, there are no studies that combine a large dataset with pattern recognition and classification techniques to study the time windows of predictions for each parameter.  

In this paper, we will analyze line-of-sight (LOS) Michelson Doppler Imager (MDI) magnetograms~\citep{scherrer1995} over solar cycle 23. We describe the complexity of each active region by extracting a large set of features of postulated importance for measuring the magnetic energy that has built up and compare these to the onset of solar flares over a range of time periods following each magnetogram. We explicitly include control data (i.e., regions that do not flare) in contrast to many studies of solely flaring regions~\citep{schrijver2007,wang2006,guo2006,meunier2004,abramenko2003,hagyard1999,zhang1994}.  We combine pattern recognition and classification techniques to determine (classify) whether a region is expected to flare; this is in contrast to many previous studies that rely merely on correlations or observations of parameters in relation to flaring activity~\citep{jing2010,ireland2008,conlon2008,georgoulis2007,schrijver2007,wang2006,guo2006,jing2006,mcateer2005,abramenko2005,meunier2004,abramenko2003,hagyard1999,zhang1994}.

Of particular interest is the characterization of local structure of AR fields.  We consider three general categories of features.  (1) Snapshots in space and time associated with increased flaring activity: (1a) total flux; (1b) magnetic gradients; (1c) neutral lines.  (2) Evolution in time: flux evolution, which can act as energy release triggers. (3) Structures at multiple size scales: wavelet analysis, which allows separation of the field into its component lengthscales. Furthermore, we will consider these features in an automated classification framework whereby we will predict flare occurrence for a series of given time windows using a combination of all above-mentioned features. 

The remainder of this paper is organized as follows.  We discuss related work in the quantification of AR complexity in Sect.~\ref{sec:ar_complexity}.  We present details of the complexity features we use in Sect.~\ref{sec:image_analysis}, including a physical motivation for each of the features and the specifics of extracting those features.  We briefly review automated classification methods and metrics for quantifying accuracy in Sect.~\ref{sec:classification}, and present results using the proposed features for classification of flare activity and discuss discriminatory features in Sect.~\ref{sec:results}.  Finally, we provide conclusions and future work in Sect.~\ref{sec:conclusions}.

\section{Analysis of active region complexity}
\label{sec:ar_complexity}
In this section, we provide a detailed overview of related work on the use of AR complexity measures for prediction and characterization of solar flares.  We focus on a listing of the specific features used in these studies (many of which we also use); accuracies, magnitudes of flares, and time windows (if published); and any significant conclusions regarding the use of photospheric complexity measures for flare prediction. We note here that it can be difficult to compare results from different papers as there are many metrics to quantify accuracy of flare prediction (several of which are defined in Sect.~\ref{sec:metrics}).  Additionally, there are differing definitions of soft x-ray flux levels which constitute flaring versus non-flaring behavior and a range of predictive time windows considered.

The location of, and the gradients along, magnetic neutral lines play a key role in many studies. \citet{ahmed2013} use machine learning to show that extensive properties connected to the neutral lines are closely connected to solar flares ($\ge$C1.0) in a 24- or 48-hour predictive time window, demonstrating flaring and non-flaring accuracies of 0.46 and 0.99, respectively. \citet{falconer2011} uses the weighted length of the strong gradient neutral line, the magnetic area, and the length of the strong-field neutral line as proxies of free magnetic energy for prediction of flares ($\ge$M1.0), coronal mass ejections, and high energy particle events over a 24-hour predictive time window. \citet{mason2010} use the total unsigned magnetic flux, primary inversion line (PIL) length, effective separation between the two polarities across the PIL, and the gradient-weighted inversion-line length (GWILL); they focus on the GWILL as the most promising measure for prediction of flares ($\ge$M1.0) for a 6-hour predictive time window, but find that it is ``not a reliable parameter.'' \citet{yuan2010}, extending upon previous work in \citet{song2009}, use the total unsigned magnetic flux, length of the strong-gradient neutral line, and the total magnetic energy dissipation for flare forecasting ($\ge$C1.0); they find weighted accuracies ranging from 0.65 to 0.86 in 24-hour forecasts of flaring. \citet{huang2010} use maximum horizontal gradient, length of the neutral line, and the number of singular points, incorporating temporal characteristics with the use of sequential supervised learning and voting by multiple classifiers; they achieve Heidke skill scores (HSS) (see Sect.~\ref{sec:hss} for a definition) of approximately 0.65 for predictive flare index $\ge$M1.0 and for a 48-hour predictive time window. \citet{welsch2009} use many properties extracted from the magnetic field and flow field to associate the properties with flaring ($\ge$C1.0) over 6- and 24-hour windows via correlation and discriminant analysis; they find the unsigned flux near strong-field polarity inversion lines to be most strongly related to flare flux, yielding climatological skill scores $\le$0.37.  \citet{song2009} use length of the strong gradient neutral line, total magnetic energy dissipation, and total unsigned magentic flux to forecast flares ($\ge$C1.0, $\ge$M1.0, and $\ge$X1.0) within a 24-hour time window; they find probabilities of detection of (0.90, 0.65, and 0.71) and false alarm rates of (0.29, 0.08, and 0.17), respectively. \citet{schrijver2007} proposes the use of the total unsigned flux `R' near high-gradient, strong-field polarity-inversion lines to characterize the electric currents in the photosphere; this parameter is found to have an increased value within a 24-hour time window for large-flare ($\ge$M1.0) producing ARs. \citet{wang2006} analyzes the relative motions of the two polarities of bipolar ARs and finds sudden change in magnetic shear following flares ($\ge$M7.9). \citet{guo2006} analyzes the effective distance (separation between flux-weighted centers of bipolar ARs), total flux, and tilt angle as compared to the Mount Wilson magnetic classification; they find that effective distance is well correlated with the Mount Wilson classes and with flaring activity ($\ge$C1.0) in $\delta$ regions. \citet{jing2006} use the mean of the spatial gradients along strong-gradient neutral lines, the length of the strong-gradient neutral lines, and the total magnetic energy dissipated in a unit layer and unit time and find positive correlation with flare ($\ge$B1.0) activity. 

Studies of local complexity across the active region have also shown some relation to solar flare activity. \citet{yu2010b} use a wavelet transform to extract multiresolution features and use the same classifier as~\citet{huang2010}, yielding an HSS of 0.77 for ARs with flare indices exceeding M1.0 and for a 48-hour predictive time window.  \citet{yu2010a} use the same parameters and extract ``sequential features'' to characterize the temporal shapes of the features; a Bayesian network achieves HSS of 0.69, again for ARs with flare indices $\ge$M1.0 and for a 48-hour predictive time window. \citet{ireland2008} use statistics of the gradient distributions along multiscale opposite polarity region separators and a Kolmogorov-Smirnov test to show that flaring ($\ge$A1.0, $\ge$M1.0) and non-flaring regions come from different gradient distributions when considered over a 6-hour time window. \citet{conlon2008} use two measures of multifractality (contributional and dimensional diversity) along with total field strength and area to postulate a relationship between multifractal properties and flaring rate. \citet{georgoulis2007} define an effective magnetic field based on connectivity and show this provides a lower limit required for M-class flares and above. \citet{mcateer2005} use the fractal dimension as determined with a modified box-counting algorithm and find that a large fractal dimension is a necessary but not sufficient condition for occurrence of large flares ($\ge$C1.0) over a 24-hour time window. \citet{abramenko2005} use structure functions to analyze multifractal properties of ARs and find that flaring regions tend to have larger degree of multifractality than do non-flaring regions. 
 
Some authors have also considered features of the photospheric magnetic field extracted from vector magnetograms. \citet{leka2003a}, \citet{leka2003b}, \citet{barnes2006}, and \citet{leka2007} develop a comprehensive list of features derived from vector magnetograms, including measures from the distribution of magnetic fields, inclination angle, spatial gradient, vertical current density, twist, current helicity, shear angles, photospheric excess magnetic energy density, and magnetic charge topology models.  They find mixed results, with some potential indicators of flare activity ($\ge$M1.0) in \citet{leka2003a}, determining combinations of variables that indicate the ability to distinguish between flaring ($\ge$M1.0) and non-flaring populations in \citet{leka2003b}, finding that coronal topology measures have better probabilities in distinguishing between flaring ($\ge$C1.0) and non-flaring regions in \citet{barnes2006}, and concluding that features based on the photospheric field have ``limited bearing on whether that region will be flare productive'' for flares $\ge$C1.0 and a 24-hour predictive time window in \citet{leka2007}.

\section{Image analysis}
\label{sec:image_analysis}
In this section we describe the three general categories of features used in this work:  (1) Snapshots in space and time, encompassing total magnetic flux, magnetic gradients, and neutral lines, (2) Evolution in time, encompassing flux evolution, and (3) Structures at multiple size scales, encompassing wavelet analysis.  For each feature category, we first discuss the theoretical background and motivation followed by discussion of the image processing methods; we break up the discussion as such to better relate the image processing and theory.

For this work, we use MDI magnetograms from solar cycle 23, including NOAA ARs 8809--10\,933, and some 260\,000 total AR cutout images.  ARs are selected based on locations specified by the Space Weather Prediction Center\footnote{\url{www.swpc.noaa.gov/ftpmenu/forecasts/warehouse.html}}.  A $300''\times300''$ window is extracted centered on these locations.  Data were cosine-corrected for line-of-sight effects, deprojected to a cylindrical equal-area mapping, and cropped to 211.5 Mm$\times$211.5 Mm (for details of these processes, see \citet{mcateer2005b}).  Additionally, magnetograms are considered only if the center of the AR is within $650''$ of disk center to mitigate projection effects and disk edge artifacts; this leaves a total of 122\,060 total AR cutout images.  It should be noted, however, that we have not implemented any correction for the saturation effect inherent in MDI data. 

\subsection{Snapshots in space and time: Total glux, gradient, and neutral line analysis}
\label{sec:gradient}
\subsubsection{Theoretical background}
At the photosphere the magnetic field is frozen into the plasma and advected by bulk plasma motions.  \citet{parker1963} showed that energy can be stored in the corona when sunspots of opposite polarity are pushed together, forming an extended current sheet above the neutral line (NL). A shear flow has a similar effect in forming a current sheet above a NL. In both cases the NL often steadily lengthens until disrupted by some instability. As such, large magnetic gradients occur across the neutral line of large spots, particularly in the vicinity of large $\delta$ spots~\citep{patty1986,zhang1994}. Over a period of hours and days, the continued concentration of opposite polarities in a relatively small area leads to strong transverse gradients~\citep{gallagher2002}. We extract features related to the overall magnetic flux present; the gradient in magnetic flux across the active region; and the size and shape of and magnetic gradient along the NL.

\subsubsection{Image processing}
We compute a total of four features related to the magnetic flux as described here and summarized in Table~\ref{tab:allfeatures}.  (1) The total unsigned magnetic flux is computed as the absolute sum of the magnetogram image.  (2) The total signed magnetic flux is computed as the sum of the magnetogram image.  (3) The total positive flux is computed as the sum of positive values of the magnetogram image.  (4) The total negative flux is computed as the sum of negative values of the magnetogram image. These features describe the total magnetic flux present in the active region, as well as the flux imbalance in the region.

\begin{figure}
 \centering
  \subfloat[Magnetogram.]{\includegraphics[width=0.45\columnwidth]{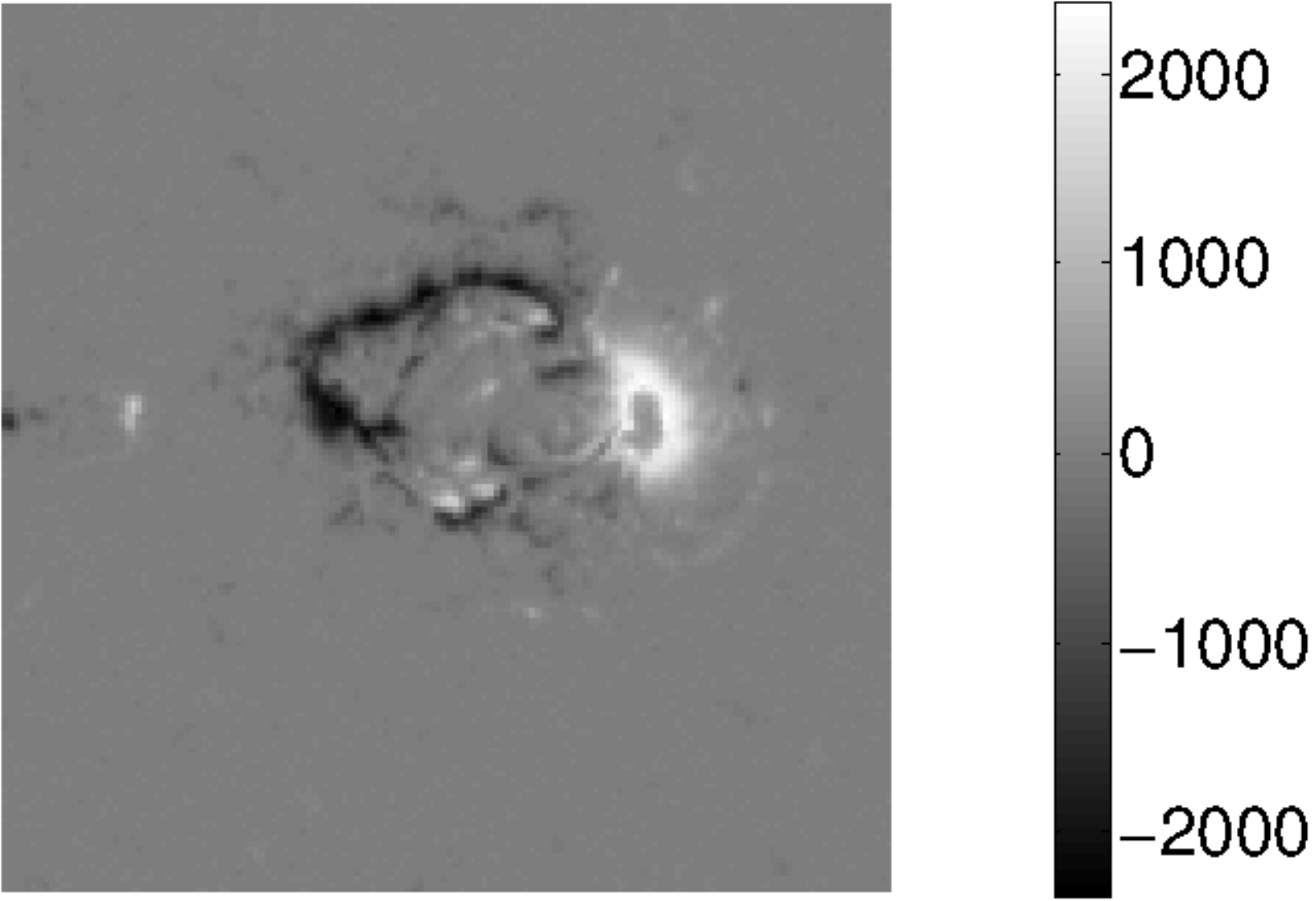}}~~~~~
  \subfloat[Gradient magnitude.]{\includegraphics[width=0.45\columnwidth]{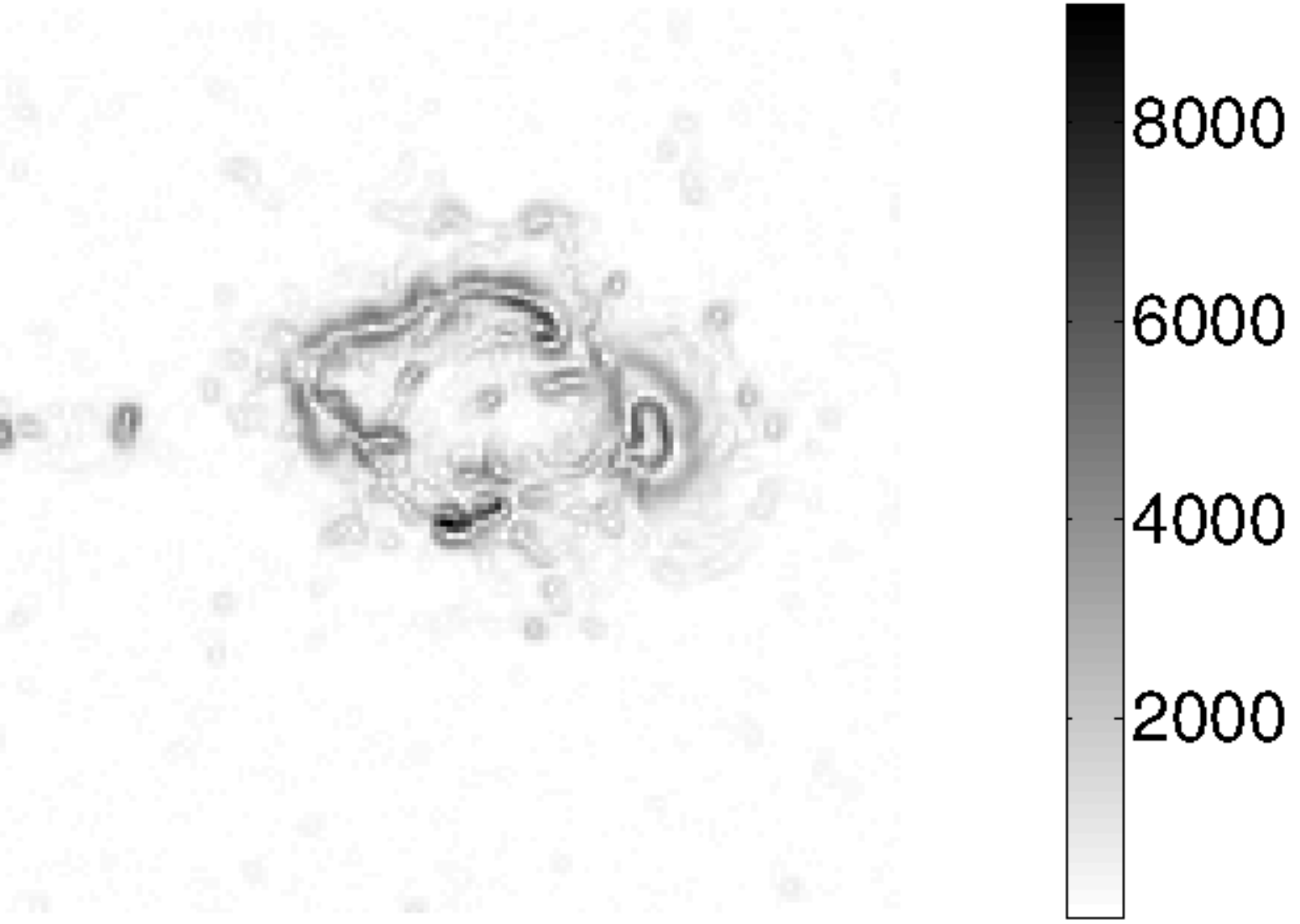}}
  \caption{Magnetogram and resultant gradient magnitude. NOAA AR \# 10\,488, 28 October 2003, 01:35.  Colorbars are in units of Gauss and each image is 211.5 Mm$\times$211.5 Mm.}
  \label{fig:gradient_result}
\end{figure}

We compute a total of 7 features related to the gradient magnitude as described here.  From the perspective of image processing, the spatial gradient is the first derivative of the image. This will highlight small specks and edges that may not be as visible in the original image. The gradient of image $f$ at coordinates $(x,y)$ is defined as the two-dimensional column vector~\citep{gonzalez2007}
\begin{equation} 
v_{f} = \left[ \begin{array}{cc} G_{x} \\ G_{y} \end{array} \right]
= \left[ \begin{array}{cc} df/dx \\ df/dy \end{array} \right]
= \left(\frac{df}{dx}\right)\hat{i} + \left(\frac{df}{dy}\right)\hat{j} ,
\end{equation}
where $G_x$ and $G_y$ are the spatial gradients in the $x$ and $y$ directions, respectively, and $\hat{i}$ and $\hat{j}$ are unit vectors in the $x$ and $y$ directions, respectively.  Gradient magnitude is defined as $|G(x,y)|=\sqrt{G_x^2+G_y^2}$. To approximate the first derivative for discretely-indexed image $f$, we use the two Sobel filters
\begin{equation} 
h_{x}=
\left[ \begin{array}{ccc}  -1 & ~~0 & ~~1 \\ -2 & ~~0 & ~~2 \\ -1 & ~~0 & ~~1 \end{array} \right],~~~
h_{y}= \left[ \begin{array}{ccc} ~~1 &~ ~2 & ~1 \\ ~~0 & ~~0 & ~~0 \\ -1 & -2 & -1 \end{array} \right] 
\end{equation}
The image is filtered (convolved) with each Sobel filter, yielding $G_x = h_x*f$ and $G_y=h_y*f$ where $*$ is the two-dimensional convolution operator.  Fig.~\ref{fig:gradient_result} shows an example of one magnetogram image and the magnitude of the spatial gradient.

The gradient magnitude is computed for each magnetogram image in our dataset. To condense this gradient information into single descriptors (features) for each image, we compute the (1) mean, (2) standard deviation, (3) maximum, (4) minimum, (5) median, (6) skewness, and (7) kurtosis of the gradients in each image as summarized in Table~\ref{tab:allfeatures}.  The gradient magnitude will be large for regions in which there are large differences in flux in close spatial proximity, and largest for opposite polarity regions with large flux in close proximity.  These seven gradient features quantify the statistics of the occurrence of large gradient magnitude.

\begin{figure}
 \centering
 \subfloat[Bipolar region: noisy zero-Gauss contour. NOAA AR \# 10\,000, 15 June 2002, 17:35, 211.5 Mm$\times$211.5 Mm.]{\includegraphics[width=0.45\columnwidth]{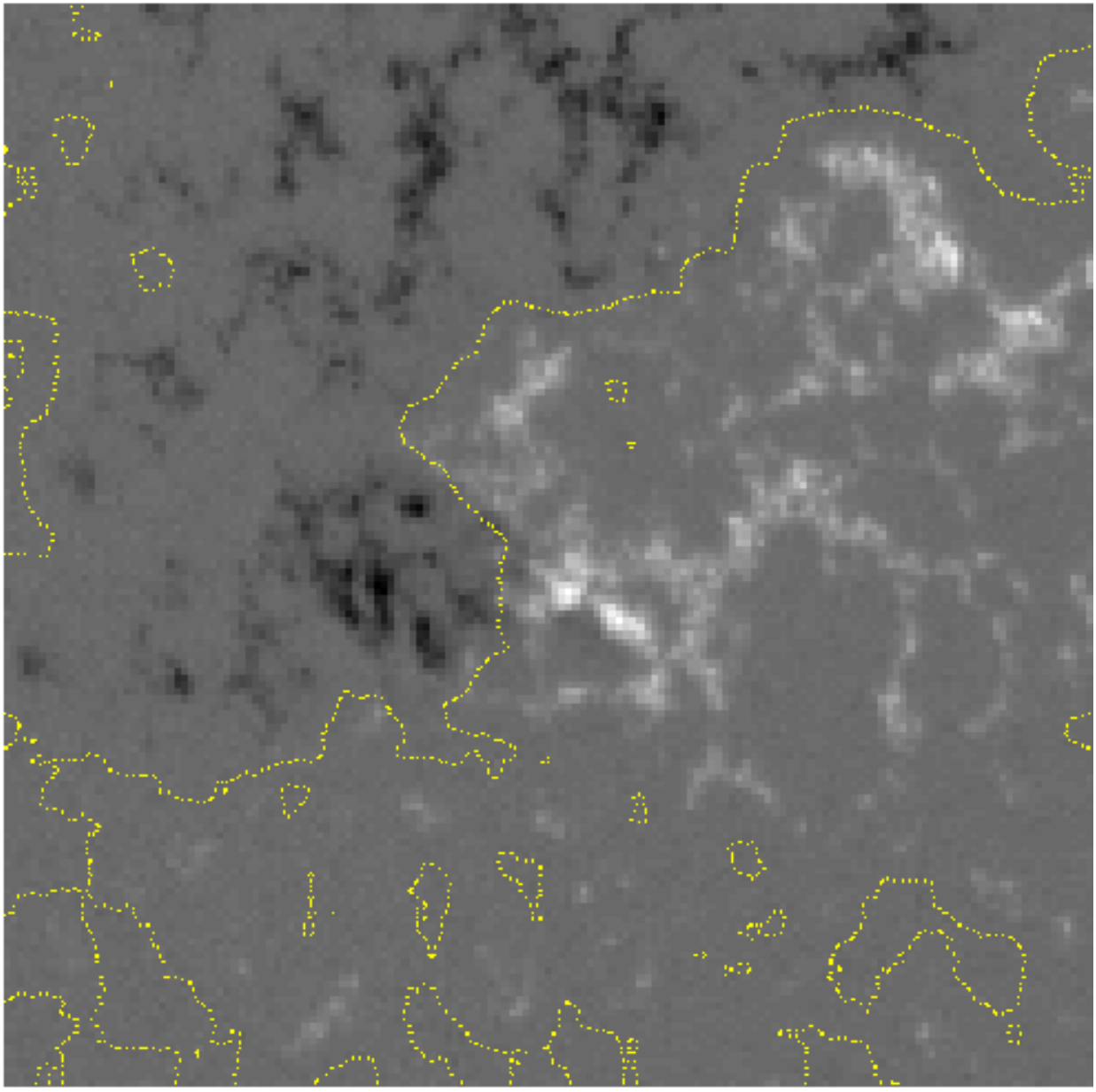}}~~~~~
 \subfloat[Bipolar region: gradient-weighted NL. NOAA AR \# 10\,000, 15 June 2002, 17:35, 211.5 Mm$\times$211.5 Mm.]{\includegraphics[width=0.45\columnwidth]{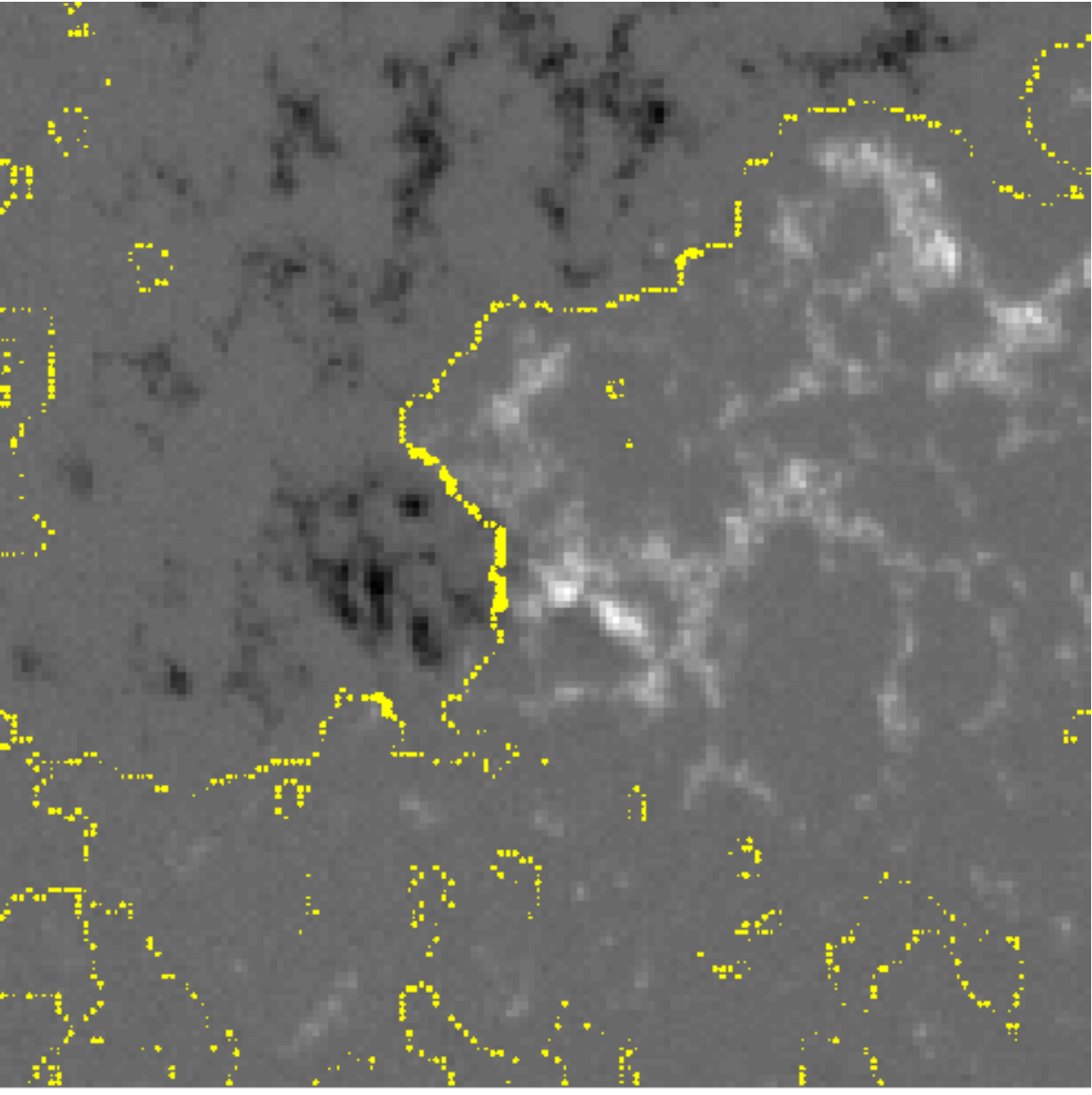}}\\
 \subfloat[Multipolar region: noisy zero-Gauss contour. NOAA AR \# 10\,488, 31 October 2003, 12:48, 211.5 Mm$\times$211.5 Mm.]{\includegraphics[width=0.45\columnwidth]{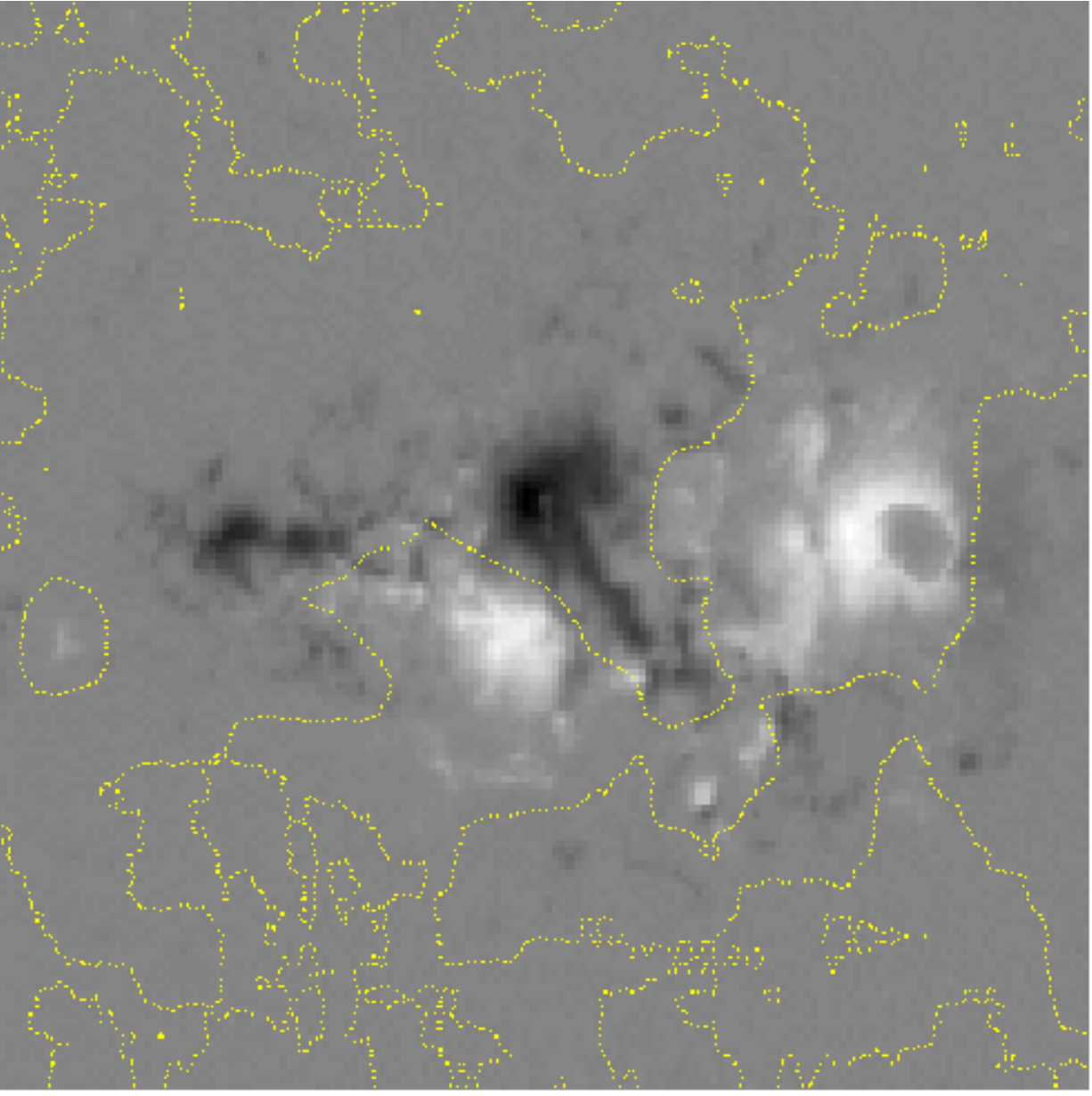}}~~~~~ 
 \subfloat[Multipolar region: gradient-weighted NL. NOAA AR \# 10\,488, 31 October 2003, 12:48, 211.5 Mm$\times$211.5 Mm.]{\includegraphics[width=0.45\columnwidth]{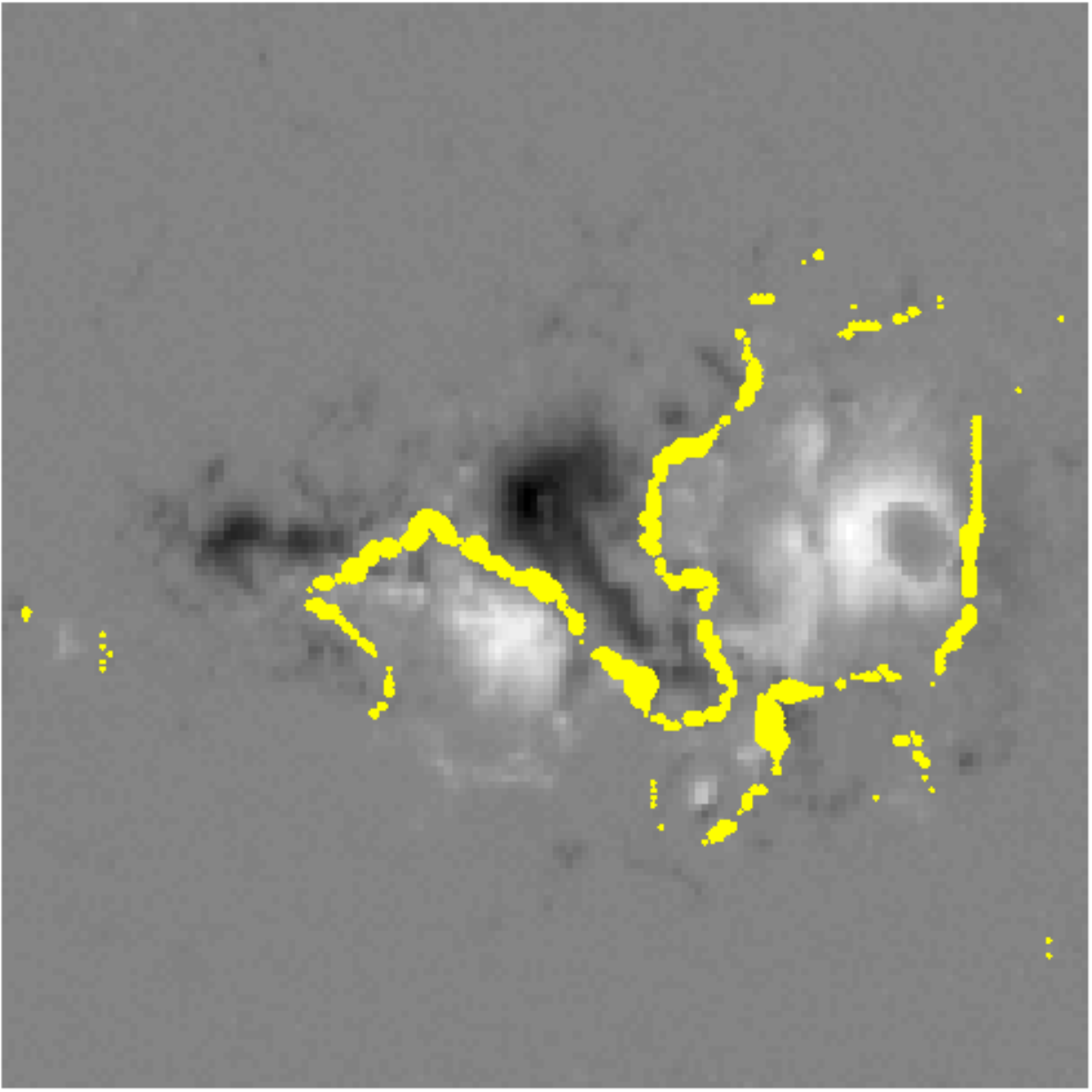}}
 \caption{Illustrations of NL analysis.  (a), (c) Noisy zero-Gauss contour before being weighted by the gradient image. We note the presence of zero-Gauss contours throughout the image, not just in the region separating the strong positive and negative flux. (b), (d) Larger yellow markers indicate the presence of a stronger gradient at that spatial location along the NL.}
 \label{fig:NL}
\end{figure}

We compute a total of 13 features related to the NL as described here.  The NL is detected in magnetogram images using the following procedure.  First, magnetogram images are smoothed using a 10$\times$10 pixel averaging filter to remove much of the statistical noise. Second, contours at the zero Gauss level of the smoothed image are used to create a NL mask.  Third, since the zero-Gauss contours will flag all pixels at the zero-Gauss level, including those with very small gradient (i.e., including those of the quiet Sun), we mask the gradient image with the NL mask.  This weights the NL to emphasize regions of the NL for which there is a large spatial gradient, indicating a transition between large positive and large negative flux. Fig.~\ref{fig:NL} illustrates all of the zero-Gauss contours for both a bipolar and multipolar region, as well as a visualization of the gradient-weighted neutral line (GWNL) for the same bipolar and multipolar ARs.  In this work, we make no distinction between the primary, high-gradient, or strong-field NL~\citep{falconer2011,yuan2010,schrijver2007,jing2006} and the NL as it exists separating opposite polarity regions. 

The NL is detected for each image in our dataset. We define a total of 13 features related to the NLs.  The first three features are defined as follows: (1) Length of the NL: The GWNL is thresholded at 20\% of the maximum value to yield a strong-gradient binary NL mask.  The NL length is determined as the sum of the pixels in this strong-gradient binary NL mask. (2) Number of fragments of the NL: Using the same thresholded GWNL, the number of fragments is defined as the number of 8-connected components in the binary NL mask image. (3) GWNL length: The gradient-weighted length of the NL, computed by summing the pixels in the GWNL image.  We note here that we define the strong-gradient NL in a relative image-by-image manner, as opposed to defining an absolute threshold for use across the dataset.  We choose a relative threshold here as we feel it is important to define ``strong'' \emph{relative} to the image in question; we will discuss this further in Sect.~\ref{sec:results}. These three features quantify different aspects of the NL length, which indicates the length of the region of most likely flux reconnection.

Additionally, ten more features are extracted based on the NL boundary curvature and NL bending energy. These features quantify the tortuousity of the NL boundary with the conjecture that a more tortuous NL indicates irregular and non-bipolar magnetic characteristics which may create more regions of probable flux reconnection. For extraction of these features, we trace the (closed) boundary of the NL in the thresholded GWNL image and compute the orientation angle of each NL boundary pixel~\citep{rodenacker2003}: 
\begin{equation}
 \theta_n = \arctan\left(\frac{y(n+1)-y(n)}{x(n+1)-x(n)}\right),~n=1,\ldots,N
\end{equation}
where $x$ and $y$ are the x- and y-coordinates of the $N$ NL boundary pixels, and by definition $x(N+1)=x(1)$ and $y(N+1)=y(1)$. The curvature angles are computed separately for each NL segment, and the mean, standard deviation, maximum, minimum, and median are computed for all curvature angles for all NL segments; we forego the computation of skewness and kurtosis since there are too few data points for accurate computation of these higher-order moments. The bending energy $B_e$ is analogous to the physical energy required to bend a rod and is computed as the normalized sum of the squared difference in curvature between subsequent boundary points~\citep{rodenacker2003}:
\begin{equation}
 B_e = \frac{1}{N}\sum_{n=1}^{N}(\theta_{n+1}-\theta_n)^2
\end{equation}
where $\theta_{N+1}=\theta_1$ by definition.  We note that the term ``energy'' in bending energy is distinct from the energy required to resist magnetic force in the AR.  We use the bending energy as a measure of the shape of the NL and as a proxy for magnetic energy built up in the AR, motivated by the fact that NLs often underlie filaments which are often the site of coronal mass ejections associated with large flares.  This measure is computed separately for each NL and the mean, standard deviation, maximum, minimum, and median are computed for the distribution of bending energy.  Table~\ref{tab:allfeatures} summarizes the 13 features extracted from NL analysis.

\subsection{Evolution in time: Flux evolution analysis}
\label{sec:EFR}
\subsubsection{Theoretical background}
Magnetic flux emergence is the origin of sunspots and ARs, and often is associated with solar eruptive events~\citep{conlon2010}. In the initial phase of AR emergence, the two opposite magnetic polarities move apart at a relatively large speed ($\sim5$ km/s) and then slow. New flux emerges continuously in the central part between the main polarities, separates and reaches the main polarities with high velocities.  Emerging flux regions (EFRs) have been shown to have significance for solar flares~\citep{tang1993} and CMEs~\citep{feynman1995, green2003}. We extract features related to flux evolution in general, and a measure of emerging flux regions.

\subsubsection{Image processing}
\begin{figure}
  \centering
 \subfloat[Magnetogram at 12:51.]{\includegraphics[width=0.45\columnwidth]{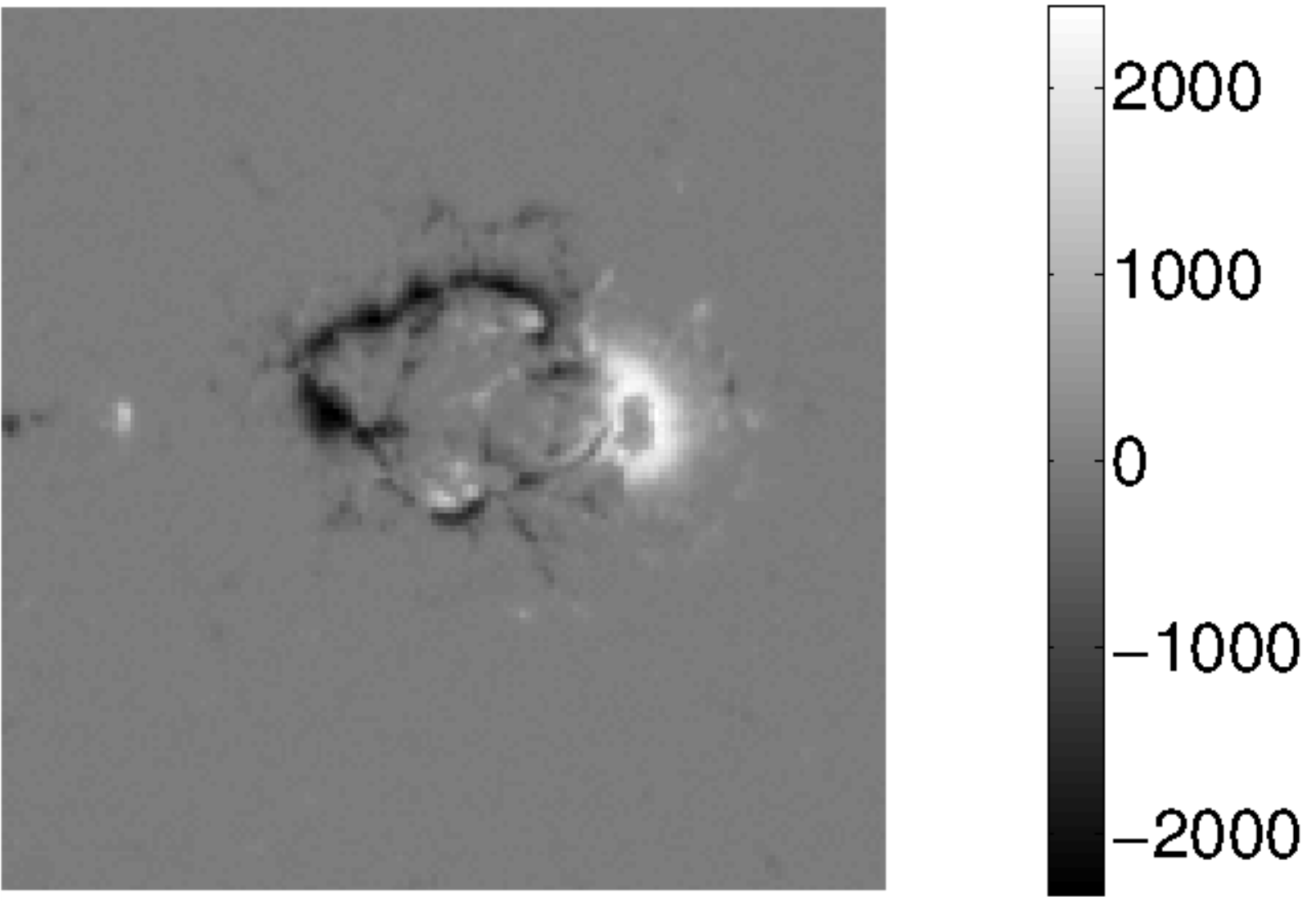}}~~~~~
 \subfloat[Magnetogram at 14:27.]{\includegraphics[width=0.45\columnwidth]{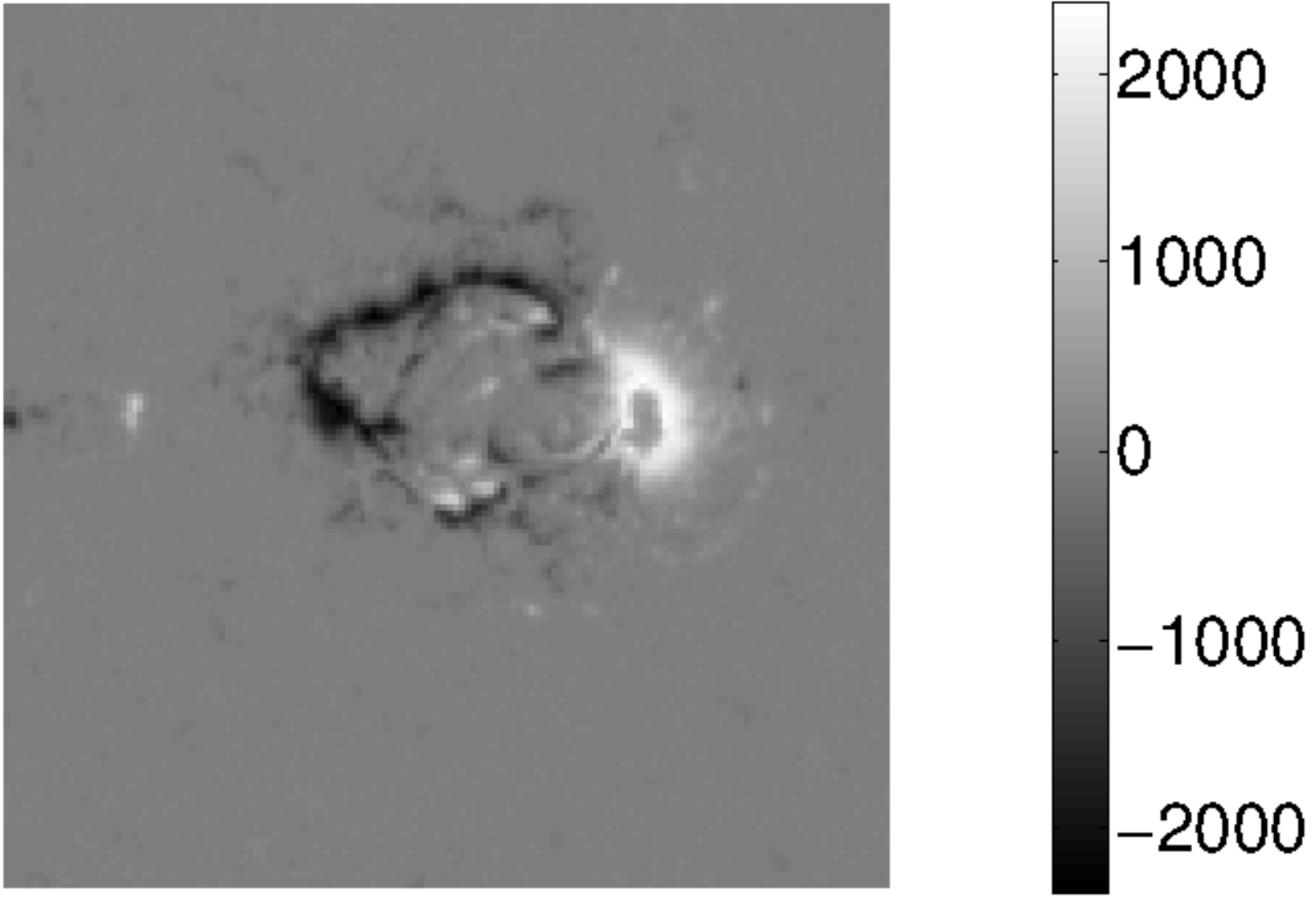}}\\
 \subfloat[Difference image.]{\includegraphics[width=0.45\columnwidth]{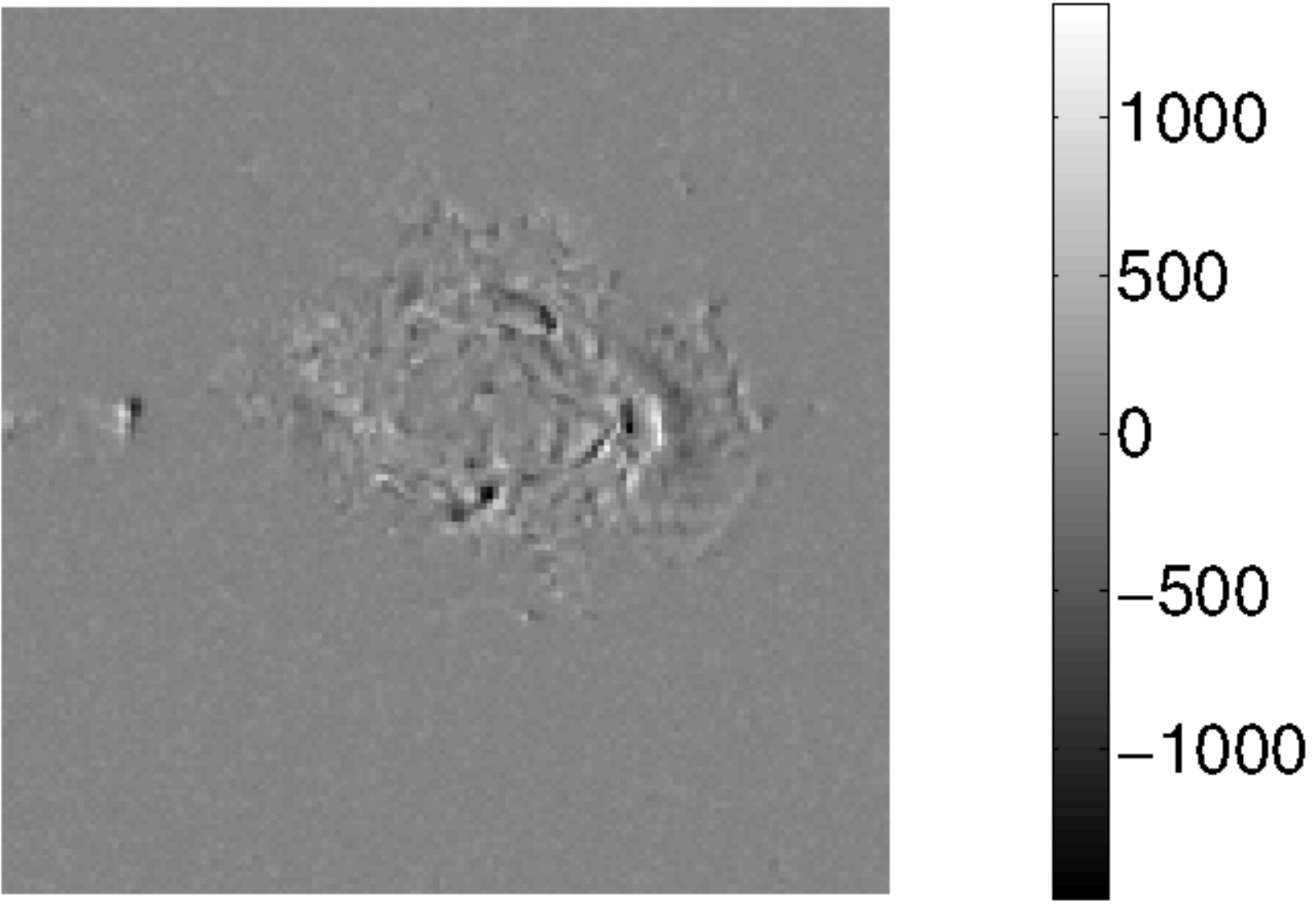}}~~~~~
 \subfloat[3$\sigma$ regions.]{\includegraphics[width=0.425\columnwidth]{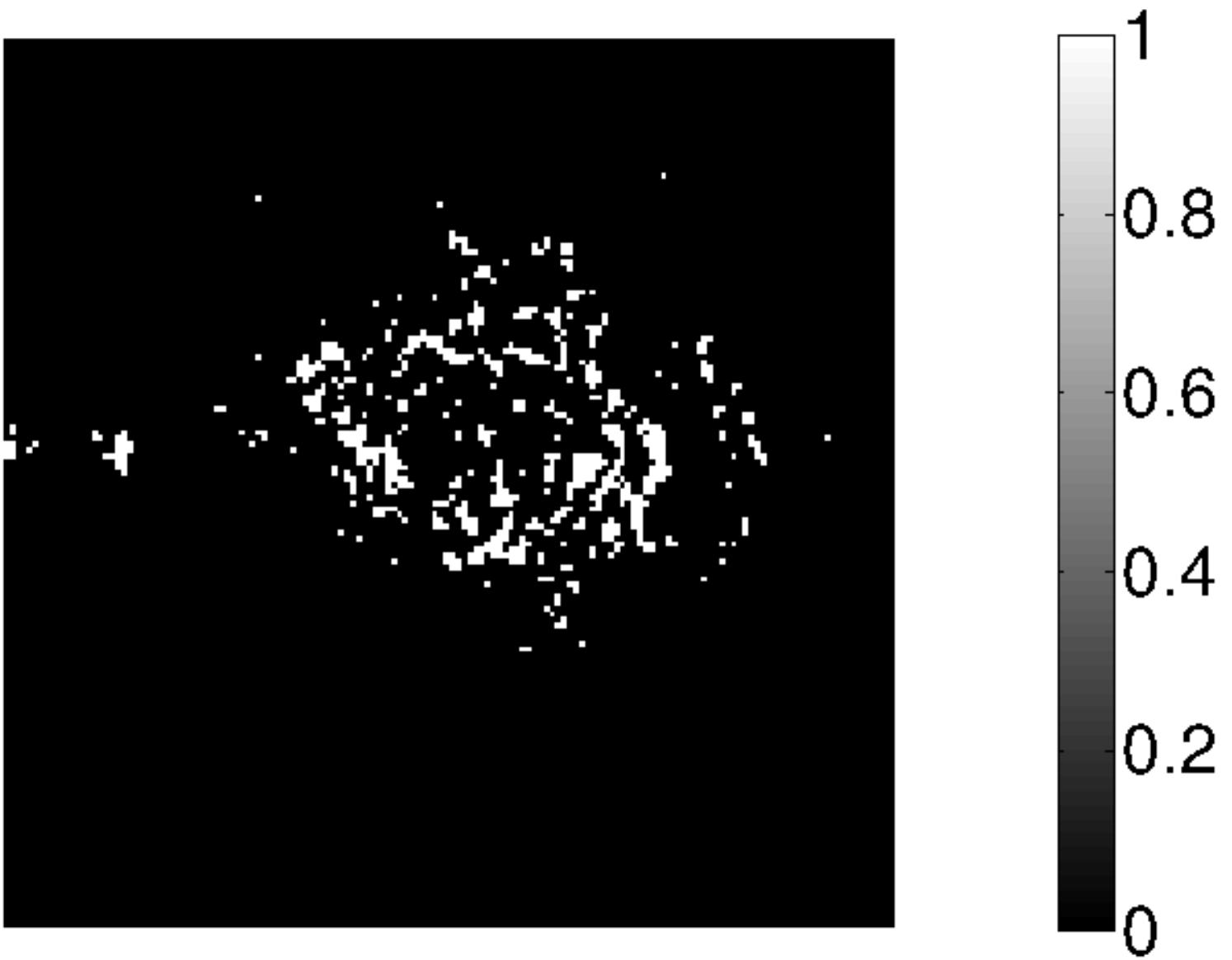}}
 \caption{Illustration of FE analysis. (a), (b) Two magnetogram images, NOAA AR \# 10\,488, 29 October 2003, 12:51 and 14:27, 211.5 Mm$\times$211.5 Mm, (c) the resultant difference image, (d) binary mask of 3$\sigma$ regions.  Colorbars for (a)-(c) are in units of Gauss, and for (d) is unitless.}
 \label{fig:EFR}
\end{figure}

Flux evolution is detected by considering difference images between two subsequent magnetograms, i.e., every two subsequent images (in one AR) are aligned and the previous magnetogram is subtracted from the following magnetogram to yield a difference image. To mitigate the effects of noise and to quantify strong changes, we identify regions in the difference image showing large deviations $(>3 \sigma)$ \emph{above} the mean difference level. Again, we choose a relative (image-by-image) threshold at the $3\sigma$ level rather than an absolute threshold to better quantify large deviations on a per-image basis; this issue will be further discussed in Sect.~\ref{sec:results}.  Fig.~\ref{fig:EFR} shows two subsequent images and the difference image along with a binary mask of the $3\sigma$ regions.

Nine features related to the difference image and flux evolution (FE) are extracted as follows and as summarized in Table~\ref{tab:allfeatures}. (1) FE sum: the sum of the difference image (the FE image), (2) FE absolute sum: the sum of the absolute value of the FE image, (3) FE gradient: the sum of the gradient image of the second magnetogram masked by the binary $3\sigma$ image, (4) FE area 3 sigma: area of the 3$\sigma$ regions. (5)-(9) FE mean, FE standard deviation, FE median, FE minimum, and FE maximum features: statistics of the FE image. It should be noted that features (1), (2), and (5)-(9) may have some contribution due to both flux emergence and submergence and features (3) and (4) explicitly characterize EFRs.

\subsection{Structures at multiple size scales: Wavelet analysis}
\label{sec:wavelet}
\subsubsection{Theoretical background}
A wavelet analysis of magnetograms discriminates spatially localized scale features such as the emergence/submergence of flux tubes.  The wavelet transform maps scale content--the power in a particular location\citep{hewett2008}.  This is essential for determining the relative influence of local magnetic features against the global properties of the AR field.  These localized magnetic features are fundamental to many flare theories and are important in developing our understanding of AR physics~\citep{ireland2008}. We extract features to quantify the structure of magnetic flux at different size scales by considering the high frequency edge content in different size scales.  Large high-frequency edge content at a particular size scale indicates the presence of flux structures at a similar size scale.  A large amount of smaller scale flux features could indicate a more complex magnetic structure with more chance for magnetic reconnection.

\subsubsection{Image processing}
The wavelet transform utilizes basis functions with compact time (spatial) support (i.e., finite in time/space).  This is in contrast to the commonly used Fourier transform whose basis functions, complex exponentials, are not compact in time (space).  Thus, the wavelet transform allows for both time (space) and frequency (wavenumber) resolution, although there is a tradeoff between the resolutions achievable simultaneously.  A variety of different basis functions can be defined, each of which has different properties in time (space) and frequency (wavenumber).  In this work, we use the Haar wavelet~\citep{gonzalez2009} and 5 levels of decomposition.

\begin{figure}
 \centering
 \includegraphics[width=0.9\columnwidth]{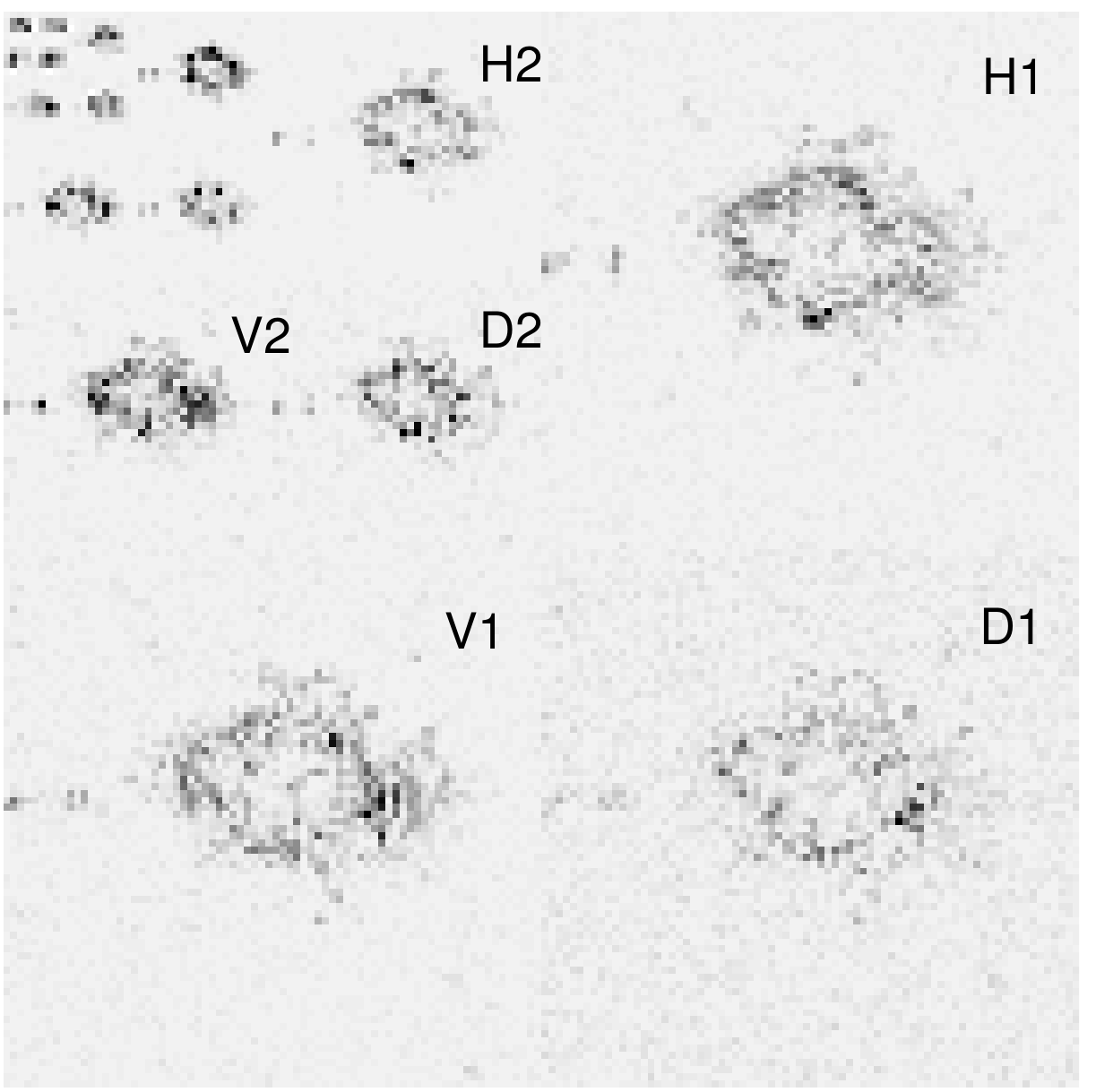}
 \caption{Five-level wavelet decomposition for NOAA AR \# 10\,488, 28 October 2003, 01:35.  The fifth-level decomposition is visualized in the uppermost left corner, with subsequently lower levels to the bottom right; as reference, the horizontal (H), vertical (V), and diagonal (D) images are labeled for the first and second level with the same structure in the subsequent levels.  The pixel values (wavelet coefficients) can be considered unitless, with darker pixels representing higher values, scaled across all levels.  The entire width of the image represents 211.5 Mm$\times$211.5 Mm; each subsequent decomposition reduces each dimension by half.}
 \label{fig:wavelet}
\end{figure}

Using the Haar transform, we can determine the resultant wavelet coefficients, yielding a low resolution image (which has been lowpass filtered and downsampled) and three highpass detail images (horizontal, vertical, and diagonal) for each level of decomposition.  Each level of decomposition involves a downsampling operation in which each of the lowpass and highpass images are reduced in resolution by a factor of 2 in each dimension.  Subsequent levels of decomposition begin with the lowpass image from the previous level.  Fig. \ref{fig:wavelet} shows the five-level decomposition for an example magnetogram image, including the lowpass and three highpass detail images for each level. The lowpass image is a decimated (lowpass filtered and downsampled) version of the magnetogram image. The three highpass detail images are a highpass filtered and downsampled version of the magnetogram image. Since highpass filters will enhance edge structure in images, the highpass detail images contain information about the edge structure at the current image resolution, indicating the presence of magnetic flux elements at a similar resolution. These three highpass detail images are used to determine the energies of each decomposition level by summing the absolute values of the wavelet coefficients (the highpass images). We sum the energies of the three highpass images together as we are interested in an orientation independent measure of edge structure.  We thus extract five energy values corresponding to each of the five levels of decomposition as summarized in Table \ref{tab:allfeatures}.

\begin{table}
\centering
\caption{Extracted Features}
\label{tab:allfeatures}
\begin{tabular}{ll}\hline\hline
\textbf{Gradient Features} & \textbf{FE Features}\\
~Gradient mean &  ~FE sum\\
~Gradient std & ~FE absolute sum\\
~Gradient median & ~FE gradient sum\\
~Gradient min & ~FE 3$\sigma$ area\\
~Gradient max & ~FE mean\\
~Gradient skewness & ~FE std\\
~Gradient kurtosis & ~FE median\\
 & ~FE min\\
\textbf{Neutral Line Features} & ~FE max\\
~NL length & \\
~NL no. fragments & \textbf{Wavelet Features}\\
~NL gradient-weighted length & ~Wavelet energy level 1\\
~NL curvature mean & ~Wavelet energy level 2\\
~NL curvature std & ~Wavelet energy level 3\\
~NL curvature median & ~Wavelet energy level 4\\
~NL curvature min & ~Wavelet energy level 5\\
~NL curvature max & \\
~NL bending energy mean & \textbf{Flux Features}\\
~NL bending energy std & ~Total unsigned flux\\
~NL bending energy median & ~Total signed flux\\
~NL bending energy min & ~Total negative flux\\
~NL bending energy max & ~Total positive flux\\\hline
\end{tabular}
\end{table}

\section{Classification}
\label{sec:classification}
In this section, we provide a brief overview of the classification method used in this work, our experimental setup, and the metrics with which we will assess performance.  In the general formulation of classification, we wish to predict some discrete target variable $t$ given some $D$-dimensional input vector $\mathbf{x}$~\citep{bishop2006}.  In the work described here, target variable $t$ corresponds to a decision that input data $\mathbf{x}$ belongs to flaring class $C_1$ or non-flaring class $C_0$.  The equation $y(\mathbf{x},\mathbf{w})$ which maps  $\mathbf{x}$ to $t$ is determined through optimization of some criterion based on training data.  

\subsection{Relevance vector machines (RVMs)}
\label{sec:RVM}
The Relevance Vector Machine (RVM)~\citep{tipping2001,tipping2003,tipping2004} is a Bayesian sparse kernel technique for regression and classification which is a probabilistic generalization of the commonly used support vector machine (SVM)~\citep{burges1998,felzenszwalb2010,melgani2004,cao2003,tong2002,hua2001,furey2000,chapelle1999,drucker1999}.  In this formulation, classification is based on the function 
\begin{equation}
 y(\mathbf{x},\mathbf{w})=\sum_{j=0}^{M-1}w_j\phi_j(\mathbf{x})=\mathbf{w}^\top\bm{\phi}(\mathbf{x})
\end{equation}
where $y$ is a function whose sign indicates the class for a given $D$-dimensional input vector $\mathbf{x}=[x_1,x_2,\ldots,x_D]^\top$; $\mathbf{w}=[w_0, w_1, \ldots, w_{M-1}]^{\top}$ is a vector of weights applied to the basis functions $\phi_j$; basis functions $\phi_j$ are some linear or non-linear function of input data $\mathbf{x}$; and $\bm{\phi}=[\phi_0,\ldots,\phi_{M-1}]^\top$~\citep{bishop2006}.  In this work, the 38-dimensional input vector $\mathbf{x}$ corresponds to the 38 features extracted for each AR image, The weight parameters $\mathbf{w}$ are chosen to optimize some criterion (the type-2 maximum likelihood in the case of RVMs), and the class indicated by $y(\mathbf{x},\mathbf{w})$ indicates the prediction of flaring or not flaring.   The weight vector $\mathbf{w}$ defines a decision boundary, a hyperplane in the multi-dimensional space spanned by $\bm{\phi}(\mathbf{x})$; data on opposite sides of this hyperplane are defined to belong to different classes.  Since the function $y(\mathbf{x},\mathbf{w})$ is linearly related to $\mathbf{w}$, the transformation $\bm{\phi}(\mathbf{x})$ allows for a non-linear decision boundary. We use the transformation $\bm{\phi}$ implicitly defined by the radial basis (Gaussian) kernel function $k(\mathbf{x},\mathbf{x}^\prime) = \bm{\phi}(\mathbf{x})^\top\bm{\phi}(\mathbf{x}^\prime) = \exp(-||\mathbf{x}-\mathbf{x}^\prime)||^2/2\sigma^2)$. 

\subsection{Experimental setup}
As output from the image analyses discussed in Sect.~\ref{sec:image_analysis}, we have one feature matrix per AR. We concatenate feature matrices for all ARs yielding feature matrix $\mathbf{X} = \left[X_{1} X_{2} ... X_{N}\right]^\top$ where $N$ is the total number of ARs.  $X_{i}$ is the feature matrix of the $i$-th AR with dimensionality $38\times n_i$, where $n_i$ is the number of images for the $i$-th AR; $n_i$ is on the order of 150 for a typical AR.  The 38 features encompass the total flux, gradient, neutral line, flux evolution, and wavelet features as discussed in Sect.~\ref{sec:image_analysis}.  Our classification is based on consideration of this feature matrix $\mathbf{X}$ one row at a time, corresponding to one AR image which is considered one data point for classification.  Within this formulation, we consider all data points (AR images) to be independent.

In addition to the feature matrix, training of supervised classifiers requires a label vector.  In this application, each element in the label vector indicates whether the AR represented by those features will flare in the next $k$ hours (`1') or not (`0'). Since the predictive time window is larger than the nominal cadence of the MDI data (96 minutes), a single flare event will be predicted (classified) independently by multiple data points (AR images). Flares are defined from the geostationary operational environmental satellite (GOES) for flare magnitude $\ge$C1.0.

\begin{figure}
  \centering
  \includegraphics[width=0.75\columnwidth]{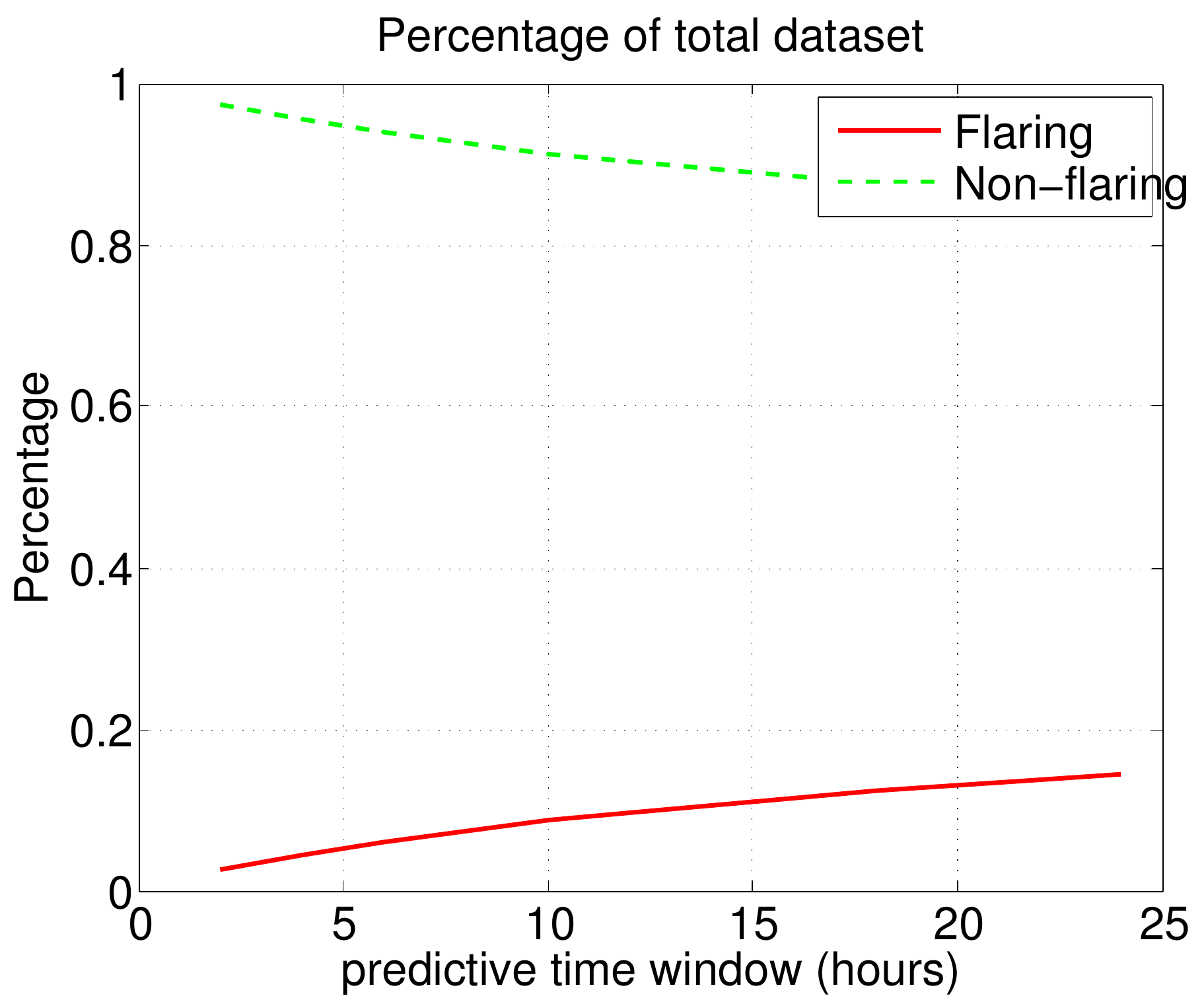}
  \caption{Percentage of total dataset comprised of flaring regions and non-flaring regions, illustrating the unbalanced nature of this dataset.}
  \label{fig:unbalanced}
\end{figure}

We randomly choose 1000 ARs to train the classifier and use a randomly subsampled balanced dataset from those 1000 ARs to find weight vector $\mathbf{w}$. A remaining 1000 ARs (and \emph{all} associated data points) are used in testing, where weight vector $\mathbf{w}$ is used to predict the class (flare or no-flare) for each data point.  Since a large majority of regions will not flare within the time window considered, classification methods can naively optimize their criterion by classifying all regions as class $C_0$ (no flare).  For example, for a 4-hour predictive time window, 4.5\% of the total dataset belongs to flaring regions and 95.5\% to non-flaring regions; this indicates that an overall accuracy of 95.5\% can be achieved simply by predicting that none of the regions will flare.  As a point of reference, we plot the percentage of the dataset comprised of flaring and non-flaring regions over all time windows in Fig.~\ref{fig:unbalanced}.

One method to alleviate the issue of unbalanced datasets is to artificially balance the dataset by subsampling one or both classes to be evenly represented.  By cross validation across many randomly balanced datasets, we can get a better idea of accuracy.  In this work, we use a 10-fold cross-validation with 500 samples each for flaring and non-flaring populations.  Thus, we randomly subsample 500 flaring data points and 500 non-flaring data points from the 1000 ARs chosen for training; this process is repeated 10 times.  Each classifier is tested on test data that has not been subsampled, consisting of 1000 ARs and some 60\,000+ data points, to yield average accuracies.  This 10-fold cross-validation is repeated for different predictive time windows in the interval [2,24] hours before flaring in a step of 2 hours.

\subsection{Metrics}
\label{sec:metrics}
The metrics we consider in this work can be derived from the basic confusion matrix (contingency table) shown in Table~\ref{tab:contingency}.  TP is the true positive (correct flare forecast), FN false negative (incorrect no-flare forecast), FP false positive (incorrect flare forecast), and TN true negative (correct no-flare forecast). 

\begin{table}[t]
\centering
\caption{Flare forecasting confusion matrix (contingency table).}
\label{tab:contingency}
\begin{tabular}{lcc}\hline\hline
         & \multicolumn{2}{c}{Forecasted}\\
Observed & Flare & No flare\\\hline
Flare        & TP & FN\\
No flare     & FP & TN\\\hline
\end{tabular}
\end{table}

\subsubsection{True positive rate (TPR) and true negative rate (TNR)}
Since flares are relatively rare events, overall classification accuracy (percentage of correctly classified data $(TP+TN)/(TP+FN+FP+TN)$) can be misleading.  As such, we present both the percentage of correctly classified flaring regions (the true positive rate or TPR, also known as the sensitivity)
\begin{equation}
 TPR = \frac{TP}{TP+FN}
\end{equation}
and the percentage of correctly classified non-flaring regions (the true negative rate or TNR, also known as the specificity)
\begin{equation}
 TNR = \frac{TN}{TN+FP}.
\end{equation}
For use in further discussions, we also include the definition of the false negative rate (FNR),
\begin{equation}
 FNR = 1-TPR = \frac{FN}{TP+FN}
\end{equation}
and the false positive rate (FPR),
\begin{equation}
 FPR = 1-TNR = \frac{FP}{TN+FP}.
\end{equation}

\subsubsection{Heidke skill score (HSS) and true skill score (TSS)}
\label{sec:hss}
The use of skill scores attempts to mitigate issues of reporting classification accuracies for unbalanced data by combining all four terms of the confusion matrix.  The Heidke skill score (HSS) and the Hanssen \& Kuipers discriminant known as the true skill score (TSS) are the most widely used in flare forecasting~\citep{bloomfield2012}. HSS is defined as:
\begin{equation}
HSS = \frac {2[(TP \times TN) - (FN \times FP)]} {(TP + FN)(FN +TN)+(TP + FP)(FP + TN)} 
\end{equation} and TSS is given by:
\begin{equation}
TSS =\frac {TP}{TP + FN} - \frac{FP}{FP + TN} 
\label{eq:TSS}
\end{equation}
We also note that $TSS = TPR-FPR = TPR-(1-TNR)$.  Only TSS is unbiased for unbalanced datasets~\citep{bloomfield2012}.

\section{Results}
\label{sec:results}
\subsection{Classification using all features}
\begin{figure*}
 \centering
  \subfloat[All 38 features.]{\includegraphics[width=0.3\textwidth]{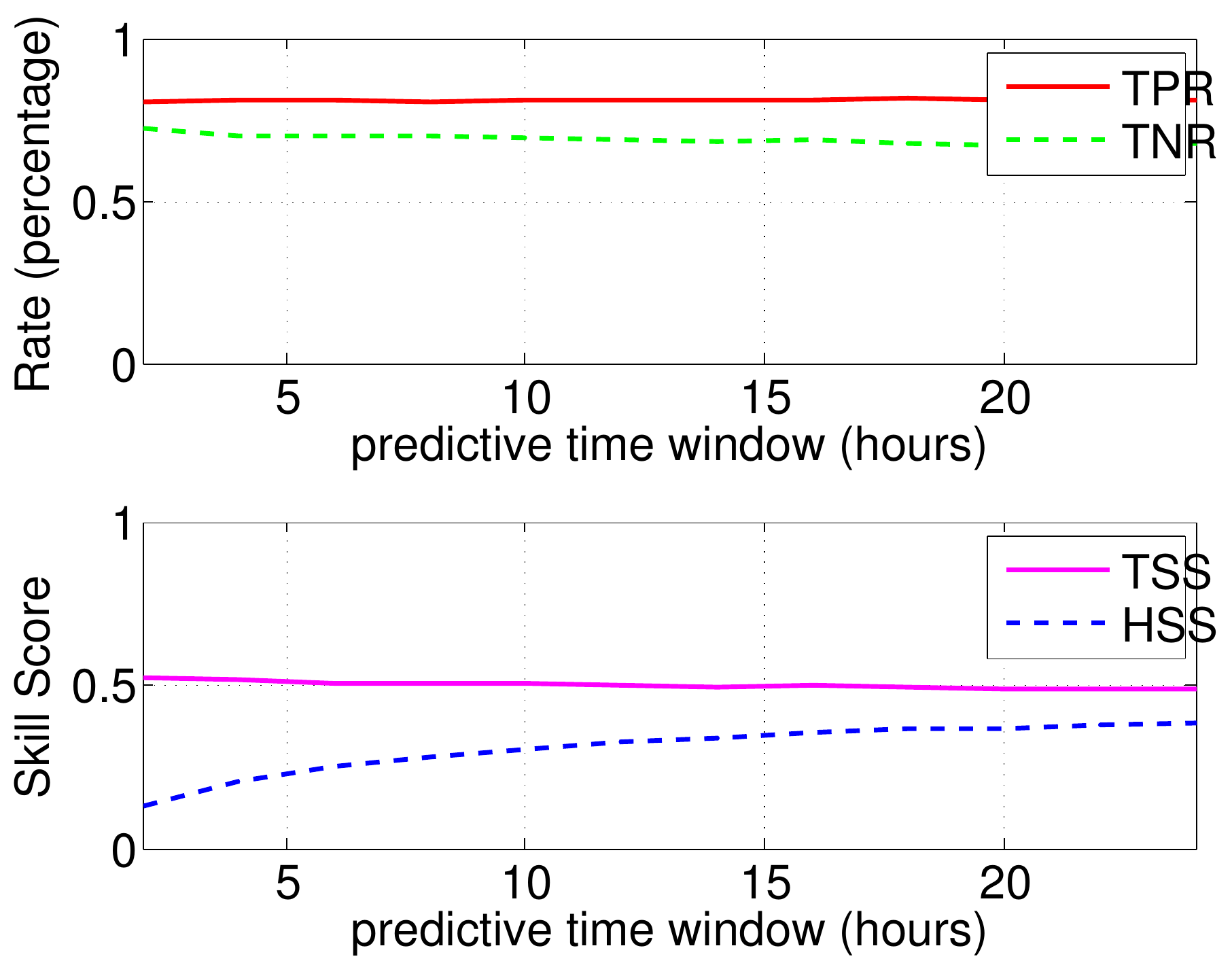}}~~~~~
  \subfloat[4 flux features.]{\includegraphics[width=0.3\textwidth]{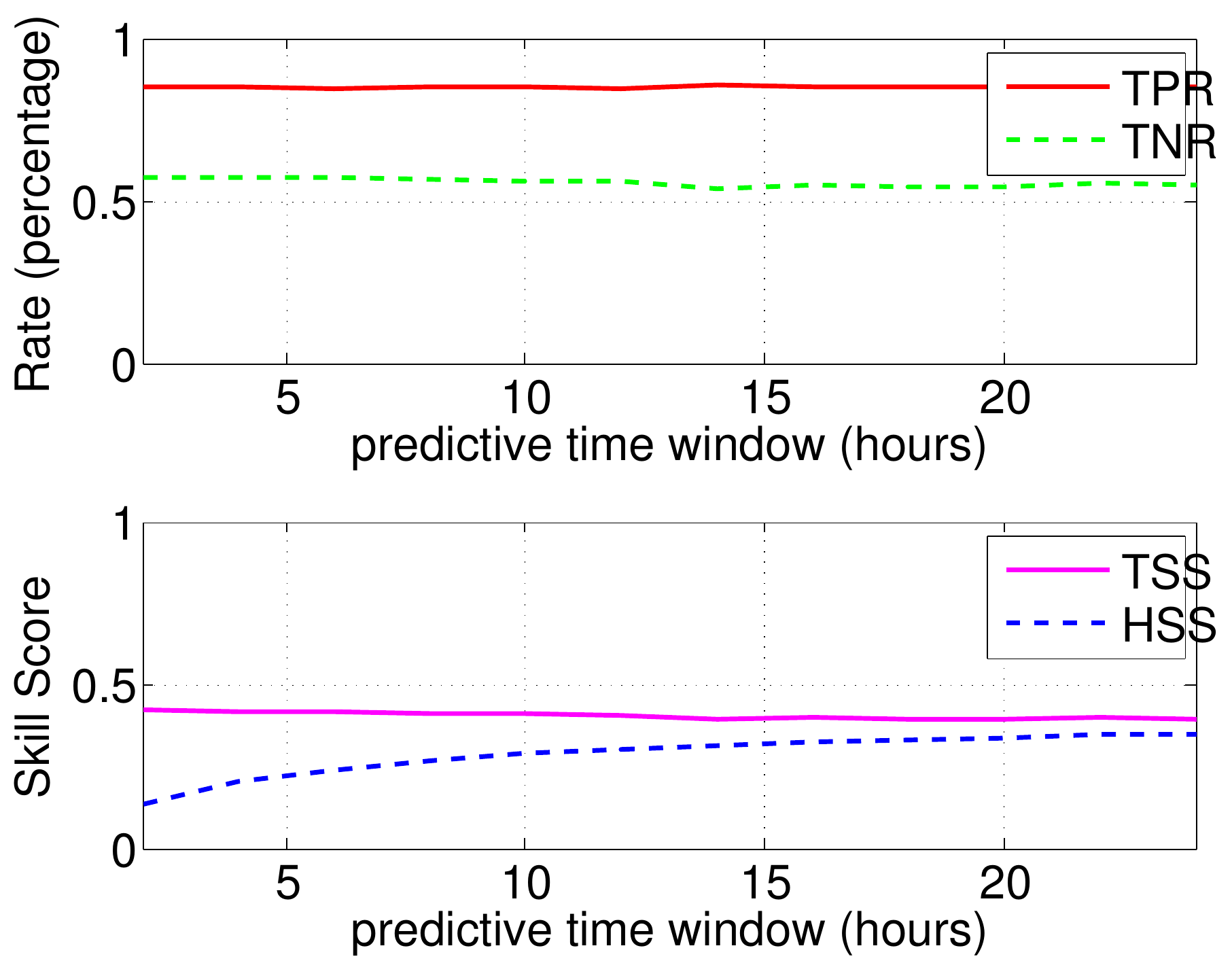}}~~~~~
  \subfloat[7 gradient features.]{\includegraphics[width=0.3\textwidth]{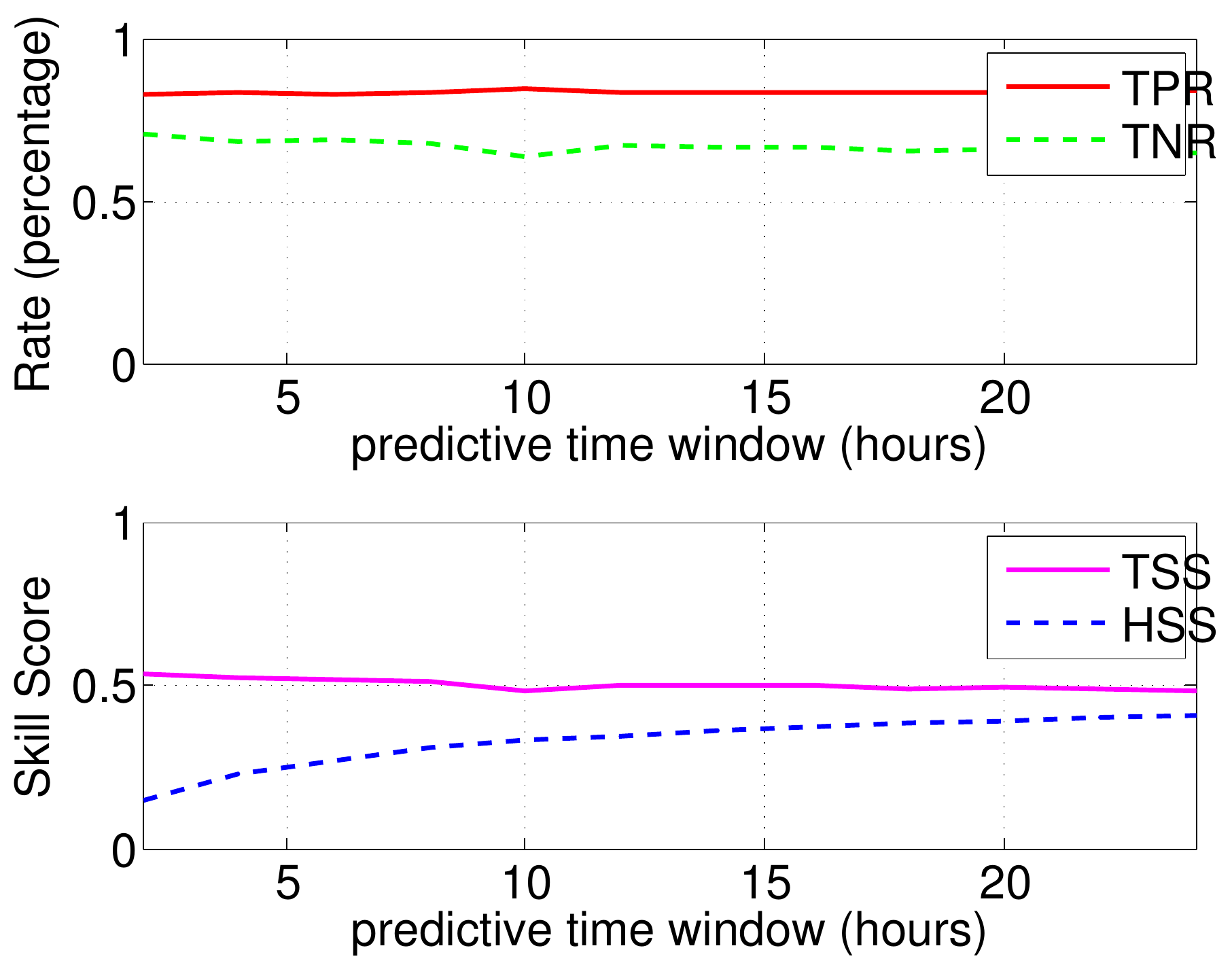}}\\
  \subfloat[13 neutral line features.]{\includegraphics[width=0.3\textwidth]{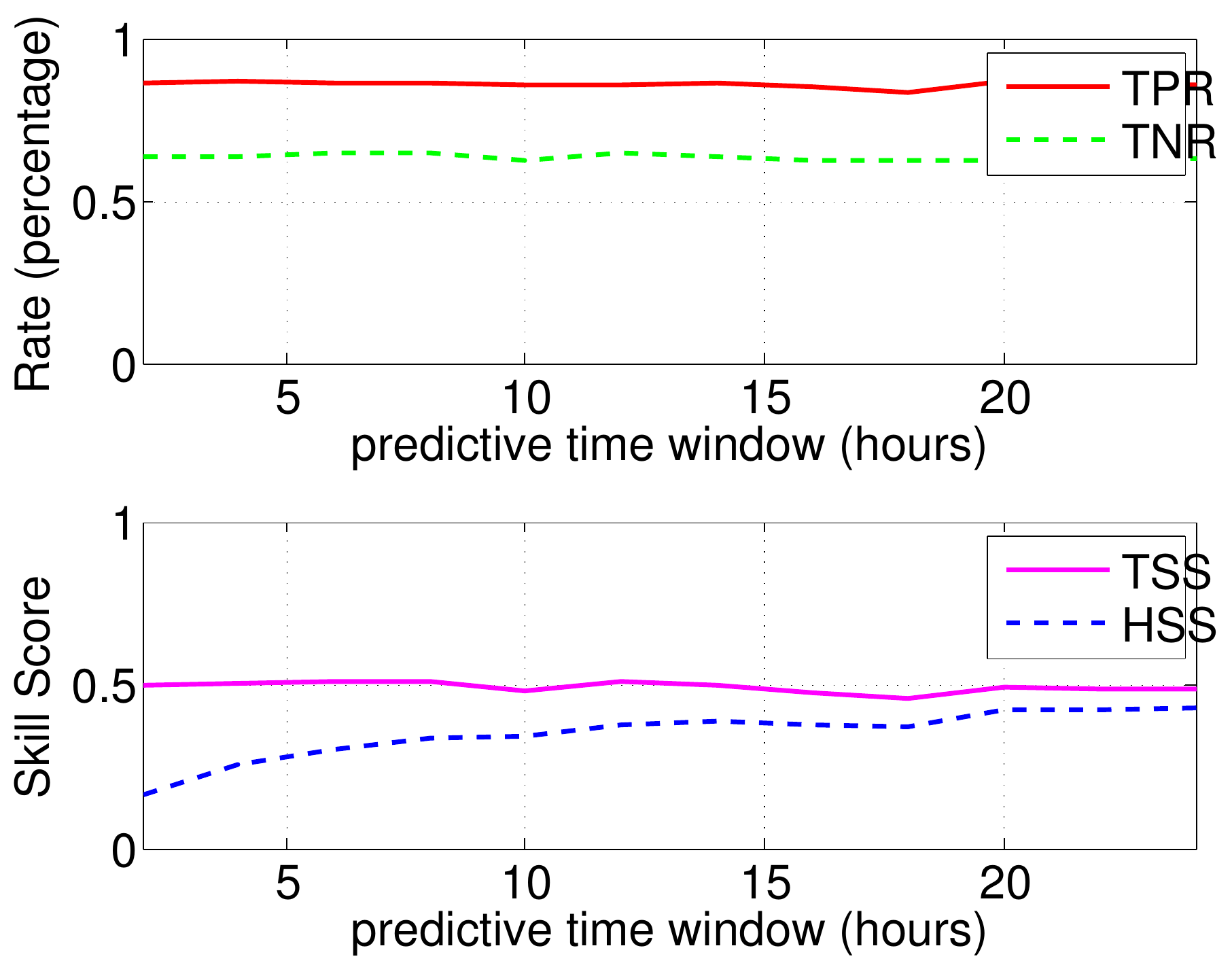}}~~~~~
  \subfloat[9 flux evolution features.]{\includegraphics[width=0.3\textwidth]{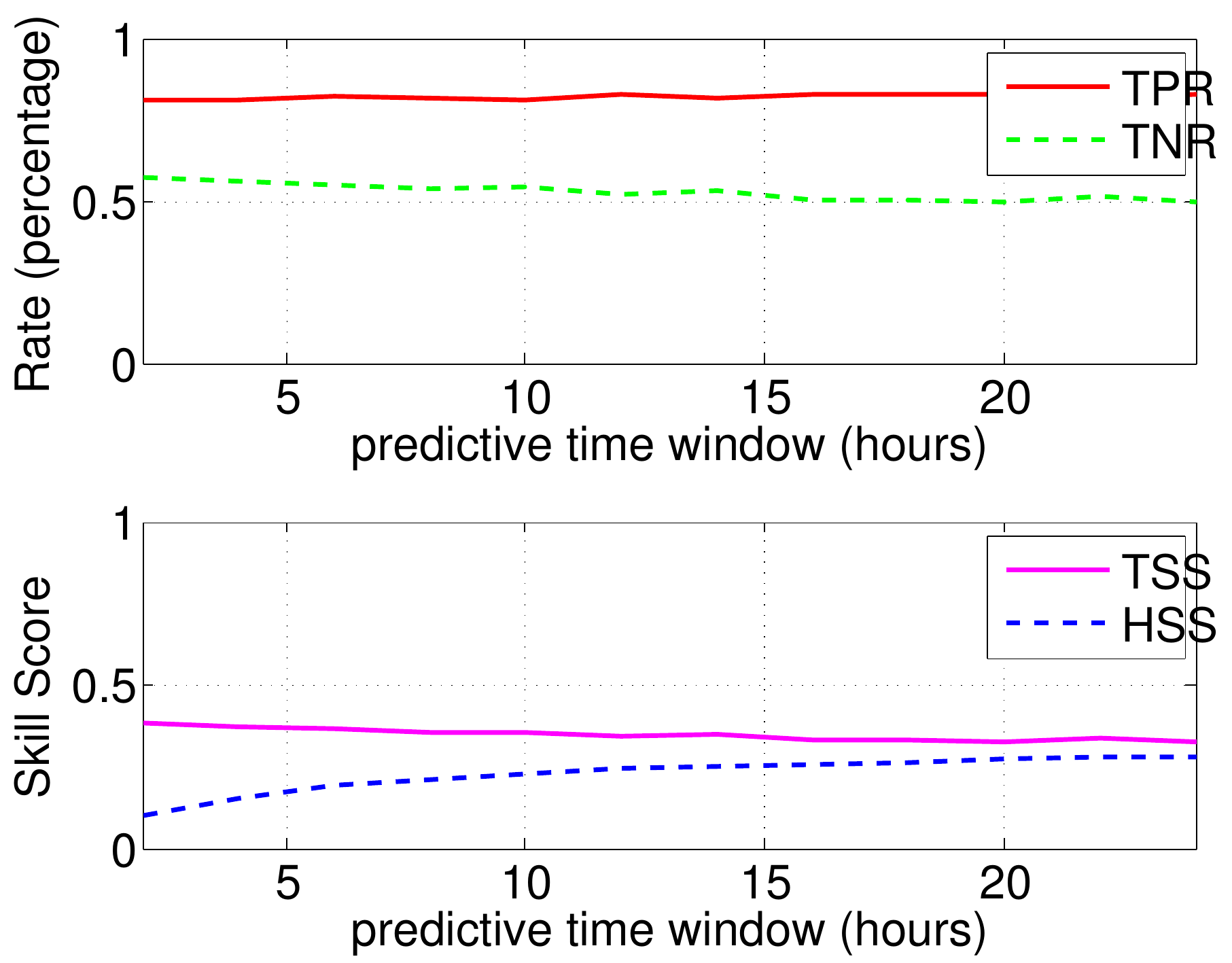}}~~~~~
  \subfloat[5 wavelet features.]{\includegraphics[width=0.3\textwidth]{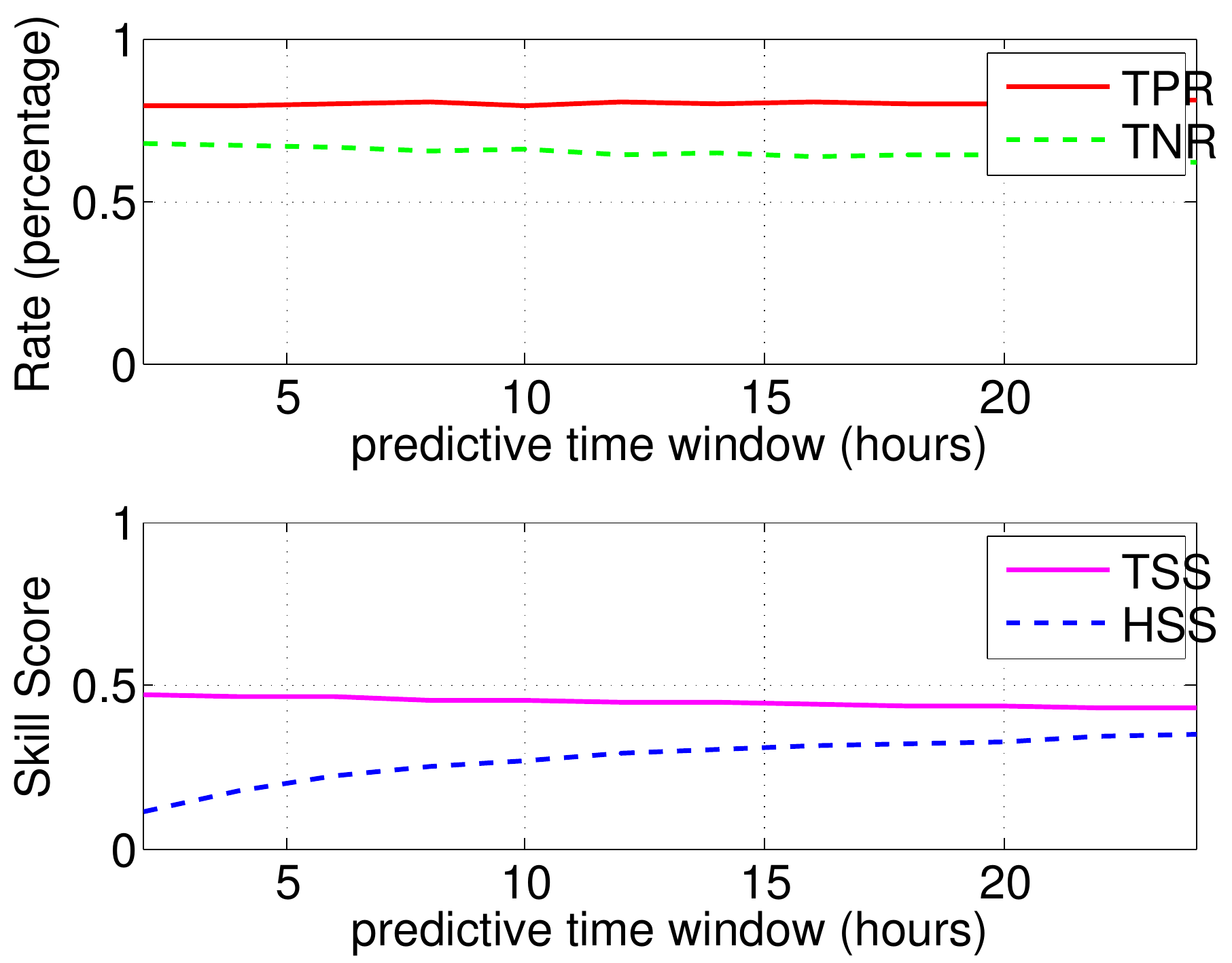}}
  \caption{TPR (\% correctly classified flaring regions), TNR (\% correctly classified non-flaring regions), HSS, and TSS for RVM classification using different feature subsets.}
  \label{fig:results}
\end{figure*}
Fig.~\ref{fig:results}(a) shows the flaring (TPR) and non-flaring (TNR) accuracies and the skill score measures (HSS and TSS) for classification of 1000 randomly chosen ARs with respect to different predictive time windows using an RVM classifier and all 38 features.  We see consistent performance across predictive time windows for both TPR ($\sim$0.8) and TNR ($\sim$0.7). We see consistent ($\sim$0.5, perhaps slightly declining) TSS performance as the predictive time window increases, while HSS increases with predictive time window.  We have also considered the standard deviation in performance across the 10 cross validation runs and found it to be 0.01--0.02 for TPR, 0.02--0.03 for TNR, 0.00-0.08 for HSS, and 0.01--0.02 for TSS.  These small standard deviations indicate that the training set is of sufficient size for generalization of the trained RVM classifier to unseen test data. It is important to note that this classification considers a region as a flaring region if it flares \emph{any} time within the predictive time window specified.  Future work will consider a regression to determine \emph{when} an AR is expected to flare; we expect that predictive time window will have a larger effect in regression analysis.

The increase in HSS with predictive time window is mainly due to its sensitivity to unbalanced datasets (as discussed in \citet{bloomfield2012} and references therein).  As predictive time window increases, the dataset becomes more balanced (see Fig.~\ref{fig:unbalanced}) while the underlying performance of the classifier (TPR and TNR) is largely unchanged. Since TPR and TNR are largely similar over the predictive time windows, the change in the four confusion matrix entries will have a much smaller effect than the change in dataset balance.

TSS may be slightly decreasing with increasing predictive time window, although it is not clear that this is a statistically significant trend. TSS may decrease if either TNR or TPR decrease; from Fig.~\ref{fig:results}, however, it appears that TNR is the most likely to be decreasing as predictive window increases. There are a variety of confounding factors which complicate the analysis of why TNR may be decreasing.  As the predictive window increases, the dataset balance is changing, and entries from the non-flare row of the confusion matrix are moving to the flare row.  If these entries were simply moving rows, we would expect TNR and TPR to move up or down in concert; we would additionally expect that TPR would move relatively more than TNR due to the imbalance of the dataset.  We do not, however, observe this in Fig.~\ref{fig:results}.  Thus, entries must be also moving between columns of the confusion matrix. This is not surprising since the RVM is now presented with different populations of data samples with which to optimize the decision boundary. In our case, it appears that entries are migrating in a fashion that has no noticeable effect on TPR, and a slight detrimental effect on TNR, indicating the non-flaring ARs are becoming more difficult to characterize as the predictive time window increases.

There are a variety of reasons that TNR could decrease, but we hypothesize that it is due to an ambiguity in predicting a non-flare. In predicting a flare, the features we use as a proxy of magnetic energy show a difference in regions that do flare versus those that do not. Once this change in features is noted, the region is more likely to flare. On the other hand, \emph{lack} of a change in feature values at a specific point in time do not provide indication that these feature values will not change at a \emph{future} point in time. This effect will be larger for larger predictive time windows. We note, however, that there is still a relatively minor degradation to either TNR or TSS over a large range of predictive time windows.

In order to determine the specific error (FN or FP) with the highest potential to improve the TSS performance, we show the confusion matrix for the 4-hour predictive time window in Table~\ref{tab:CM_example}. We note that the TPR term contributes positively to the TSS at a rate of 0.70 while the FPR contributes negatively by 0.19.  While increasing either the TPR or decreasing the FPR (increasing the TNR) can improve the TSS, there will be more advantage to improving the TPR since that is the variable with the most room for improvement.  We note, however, that the consequence of each of the two errors FP and FN may be significantly different, which may justify more focus to improving either TPR or FPR, independent of the room for improvement.  Of course, since all four measures in the confusion matrix are inherently related, it may be difficult to improve TPR without negatively affecting FPR.  We will discuss several potential modifications to the classification process that may improve TSS in Sect.~\ref{sec:conclusions}.

\begin{table}
\centering
\caption{Flare forecasting confusion matrix (contingency table) for all features, 4-hour predictive time window.}
\label{tab:CM_example}
\begin{tabular}{lcc}\hline\hline
         & \multicolumn{2}{c}{Forecasted}\\
Observed & Flare & No flare\\\hline
Flare        & TP=2269 & FN=956\\
No flare     & FP=11\,077 & TN=47\,671\\\hline
\end{tabular}
\end{table}

\subsection{Classification using feature subsets}
We consider the classification performance using subsets of features by training and testing an RVM on a subset of features.  For example, we train and test using the 4 flux features and achieve performance as shown in Fig.~\ref{fig:results}(b). In a similar manner, classification using other feature subsets are shown in Fig.~\ref{fig:results}(c)--(f).  Our goal here is to determine which subsets may have better accuracy and to allow for future work in postulating the physical relation between features and AR flaring. We note similar performance for TPR, between 0.80 and 0.85.  There are significant differences, however, for the TNR performance of different feature subsets, ranging from 0.45 to 0.69.  We find standard deviation in performance across the 10 cross validation runs to be 0.01--0.06 for TPR, 0.01--0.11 for TNR, 0.00-0.09 for HSS, and 0.00--0.07 for TSS.  These ranges in standard deviation consider the range across all of the feature subsets.  In general, the poorer performing feature subsets tend to display a larger standard deviation.

These differences in performance can be considered simultaneously in the TSS plots (since TSS is linearly related to both TPR and TNR).  In particular, we note that the gradient features yield the highest TSS, while the FE features yield the lowest.  Indeed, the gradient features alone yield performance very similar to that of all the features combined. We note a similar trend in the feature subset results to the 38-feature results in that performance is largely similar across the range of predictive time windows.

\subsection{Classification using individual features}
As a further study of the discriminatory potential of specific features, we consider the classification performance using single features as input to the RVM.  As in the experiments with feature subsets, we train and test an RVM on a single feature.  The performance over features and predictive time windows is summarized in Table~\ref{tab:ranking} where we show the top five features ranked according to their TSS.  We find much consistency in the top ranked features, with the standard deviation of the spatial gradient being the top-ranked feature for all but the 2-hour predictive time window.  Other commonly occurring features include the maximum gradient, mean gradient, and the various wavelet energies.  We find no significant differences in discriminatory features across the different predictive time windows.  

\begin{table*}
\centering
\caption{Top 5 features and TSS values (in parenthesis) for RVM classification using individual features. MF are magnetic flux features, G gradient features, NL neutral line features, FE flux evolution features, and W wavelet features.}
\label{tab:ranking}
\begin{tabular}{lllllll}\hline\hline
Rank & 2 hours     & 4 hours        & 6 hours        & 8 hours        & 10 hours       & 12 hours\\\hline
1 & G max (0.47)   & G std (0.46)   & G std (0.47)   & G std (0.46)   & G std (0.45)   & G std (0.45)\\
  &                & W L4 (0.46)    &                &                & W L4 (0.45     & \\
2 & G std (0.46)   & G max (0.45)   & W L4 (0.46)    & W L4 (0.45)    & W L3 (0.44)    & W L4 (0.44)\\
  & W L3 (0.46)    & W L3 (0.45)    &                & W L3 (0.45)    & W L5 (0.44)    & W L3 (0.44)\\
  & W L4 (0.46)    & W L5 (0.45)    &                &                &                &\\
3 & W L5 (0.45)    & G mean (0.44)  & G max (0.45)   & G max (0.44)   & G mean (0.43)  & W L5 (0.43)\\
  & G mean (0.45)  & W L2 (0.44)    & W L3 (0.45)    & W L5 (0.44)    & W L2 (0.43)    & G mean (0.43)\\
  & W L2 (0.45)    &                &                &                & G max (0.43    & W L2 (0.43)\\
4 & FE std (0.34)  & MF Neg (0.37)  & W L2 (0.44)    & G mean (0.43)  & FE std (0.31)  & G max (0.41)\\
  &                &                & G mean (0.44)  & W L2 (0.43)    &                & \\
  &                &                & W L5 (0.44)    &                &                & \\
5 & MF neg (0.31)  & FE std (0.32)  & FE std (0.32)  & MF Neg (0.34)  & MF neg (0.29)  & FE std (0.30)\\\hline\hline
Rank & 14 hours    & 16 hours       & 18 hours       & 20 hours       & 22 hours       & 24 hours\\\hline
1 & G std (0.45)   & G std (0.45)   & G std (0.45)   & G std (0.45)   & G std (0.44)   & G std (0.45)\\
2 & W L3 (0.44)    & W L4 (0.43)    & W L4 (0.43)    & W L4 (0.43)    & W L4 (0.43)    & W L4 (0.43)\\
  & W L4 (0.44)    & W L3 (0.43)    & W L3 (0.43)    & W L3 (0.43)    & W L3 (0.43)    & \\
  &                & W L5 (0.43)    &                &                &                & \\
  &                & W L2 (0.43)    &                &                &                & \\
  &                & G mean (0.43)  &                &                &                & \\
3 & W L2 (0.43)    & G max (0.41)   & W L5 (0.42)    & W L5 (0.42)    & W L5 (0.42)    & W L3 (0.42)\\
  & W L5 (0.43)    &                & G mean (0.42)  & W L2 (0.42)    & W L2 (0.42)    & W L5 (0.42)\\
  &                &                & W L2 (0.42)    & G mean (0.42)  & G mean (0.42)  & W L2 (0.42)\\
4 & G mean (0.42)  & FE std (0.29)  & G max (0.41)   & G max (0.41)   & G max (0.39)   & G mean (0.41)\\
5 & G max (0.41)   & MF neg (0.25)  & FE std (0.29)  & FE std (0.30)  & FE  std (0.29) & G max (0.39)\\\hline
\end{tabular}
\end{table*}

We make three observations regarding the performance of individual features.  First, we note that all features with TSS$>$0.40 are either gradient or wavelet features.  This indicates that the most discriminatory features come from either gradient analysis or wavelet analysis which, at a basic level, quantify edge strengths in the magnetogram.  Second, we note that the various statistics of the gradient image and the energies of the various size scales of the wavelet analysis provide largely the same discriminatory potential. Third, we note that it is interesting that the gradient standard deviation alone can achieve a TSS close to that of the classification using all features.  It is important to note, however, that the individual feature performance does not indicate the discriminatory potential for a feature when \emph{combined} with other features (as in the feature subset plots in Fig.~\ref{fig:results}). In future work, we will consider the use of optimal subsets of features, as further discussed in Sect.~\ref{sec:conclusions}.

\subsection{Relative versus absolute thresholds}
In this work, we chose to use relative thresholds for segmenting the NL (20\% of the maximum value of the gradient-weighted NL) and the FE 3$\sigma$ regions (3$\sigma$ above the mean value  of the different image).  To study the effect of using relative versus absolute thresholds for computation of these features, we ran the classification simulations with feature computed using two different absolute thresholds.  In the first case, we chose an absolute threshold for both the NL and FE features to be the mean of the relative threshold across the entire dataset, resulting in thresholds of 384 G for the NL and 54 G for the 3$\sigma$ regions.  In the second case, we used a threshold for both the NL and FE features of 50G, a common threshold used in the literature for ``strong'' flux.

In the first case (384 G for NL and 54 G for FE), we see a slight increase in performance for the NL features which also positively affected the results for all features.  In particular, we note an increase in TSS of approximately 0.03 across the predictive time windows for both the NL features alone and for the 38-feature results.  The use of this absolute threshold had no noticeable effect on the TSS of the FE features.  In the second case (50 G for both NL and FE), we see a slight decrease in performance for the NL features of approximately 0.03 in TSS.  The decreased performance of the NL features did not affect the overall 38-feature results.  The use of this absolute threshold had no noticeable effect on the TSS of the FE features. It is unclear whether any of these differences in performance are statistically significant, as they are only slightly larger than the 0.02 standard deviation in performance measured across the 10 cross validation runs.  Future work will consider in more detail the effects of relative versus absolute thresholds.

\subsection{Comparison to related work}
\begin{table*}
\centering
\caption{Comparison to related flare prediction methods.}
\label{tab:compare}
\begin{tabular}{lllllllllll}\hline 
Reference & ARs                  & Images              & Flares\tablefootmark{a}        & Magnitude                  & Window (hr) & Temporal & TPR            & TNR                         & TSS                    & HSS\\\hline

Ours      & 2124                 & 122\,060            & 3432--19\,086\tablefootmark{b} & $\ge$C1.0                  & 2--24              
            & No       & 0.81           & 0.70                        & 0.51                   & 0.39\\\hline

1         & N/A\tablefootmark{c} & ?\tablefootmark{d}  & 8498                           & $\ge$C1.0                  & 24          
            & No       & 0.46\tablefootmark{e} & 0.99\tablefootmark{e} & 0.45\tablefootmark{f} & 0.54\tablefootmark{e} \\\hline

2         & 230                  & ?\tablefootmark{d}  & 167                            & $\ge$C1.0                  & 24              
            & No       & 0.33\tablefootmark{g} & 0.92\tablefootmark{g} & 0.25\tablefootmark{g} & 0.29\tablefootmark{g}\\\hline

3         & 870                  & 48\,344             & 8612                           & $\ge$M1.0\tablefootmark{h} & 48           
            & Yes      & 0.90\tablefootmark{i} & 0.88\tablefootmark{i} & 0.78\tablefootmark{f} & 0.66\tablefootmark{i}\\\hline

4         & ?\tablefootmark{d}   & 31\,164             & 8510                           & $\ge$M1.0\tablefootmark{h} & 48           
            & Yes      & 0.85\tablefootmark{j} & 0.88\tablefootmark{j} & 0.73\tablefootmark{f} & 0.69\tablefootmark{j}\\\hline

5         & 1010                 & 55\,582             & 9801                           & $\ge$M1.0\tablefootmark{h} & 48           
            & Yes      & 0.95\tablefootmark{k} & 0.92\tablefootmark{k} & 0.87\tablefootmark{f} & 0.77\tablefootmark{k}\\\hline
 
6         & 46                   & 2708                & 119\,228\tablefootmark{b}      & $\ge$C1.0                  & 6, 24       
            & No       & 0.49\tablefootmark{l} & 0.99\tablefootmark{l} & 0.47\tablefootmark{l} & 0.51\tablefootmark{l}\\\hline

7         & ?\tablefootmark{d}   & 55                  & 54                             & $\ge$C1.0                  & 24          
            & No       & 0.75\tablefootmark{m} & 0.93\tablefootmark{m} & 0.68\tablefootmark{m} & 0.69\tablefootmark{m}\\\hline
\end{tabular}
\tablebib{(1) \citet{ahmed2013}; (2) \citet{yuan2010}; (3) \citet{huang2010}; (4) \citet{yu2010b}; (5) \citet{yu2010a}; (6) \citet{welsch2009}; (7) \citet{song2009}.}
\tablefoot{
\tablefoottext{a}{Number of data points associated with a flare; multiple data points may include the same flare within the time window.}
\tablefoottext{b}{The range of values is due to the range in time windows.}
\tablefoottext{c}{Magnetic features are considered rather than ARs.}
\tablefoottext{d}{This information was not readily apparent from the paper.}
\tablefoottext{e}{From Table 5 in \citet{ahmed2013}.}
\tablefoottext{f}{Computed from TPR and TNR.}
\tablefoottext{g}{Compiled from Figures 3-6 in \citet{yuan2010}.}
\tablefoottext{h}{This magnitude specifies the total flare importance index.}
\tablefoottext{i}{From Figure 5 in \citet{huang2010}.}
\tablefoottext{j}{From Table 5, BN\_F column in \citet{yu2010b}.}
\tablefoottext{k}{From Table 5, MODWT\_DB2\_Red column in \citet{yu2010a}.}
\tablefoottext{l}{Computed from Table 4, 24N FLCT in \citet{welsch2009}.}
\tablefoottext{m}{Computed from Table 8, Model (7) in \citet{song2009}.}
}
\end{table*}
We now discuss our results in light of results published in related work, particularly those with quantitative metrics of performance~\citep{ahmed2013,yuan2010,huang2010,yu2010a,yu2010b,welsch2009,song2009}.  As mentioned in Sect.~\ref{sec:ar_complexity}, use of different datasets, accuracy metrics, flare magnitudes, and time windows can complicate direct comparision of results. In this discussion, we highlight similarities and differences in the methods as well as datasets, metrics, magnitudes, and time windows. Additionally, we summarize some key characteristics of the dataset and performance in Table~\ref{tab:compare}.

We discuss here some of the similarities and differences in the datasets for the aforementioned work.  First, we note that our dataset, at 2124 active regions and 122,060 total images is over twice the size of the largest dataset considered in other work besides \citet{ahmed2013}, which considers magnetic features rather than NOAA active regions.  Second, we note that the magnitude considered to constitute a flare is $\ge$C1.0 for our work, \citet{ahmed2013}, \citet{yuan2010}, \citet{welsch2009}, and \citet{song2009}; other work considers regions with a total flare \emph{importance index} of $\ge$M1.0~\citep{huang2010,yu2010a,yu2010b}.  Third, we consider a range of predictive time windows from 2 to 24 hours; other work considers 6 hours~\citep{welsch2009}, 24 hours~\citep{ahmed2013,yuan2010,welsch2009,song2009}, or 48 hours~\citep{ahmed2013,huang2010,yu2010a,yu2010b}. Fourth, we note that some researchers have begun using temporal information for flare prediction~\citep{huang2010,yu2010a,yu2010b}.

Compared to those works that do not use temporal information~\citep{ahmed2013,yuan2010,welsch2009,song2009}, we find our method to have a higher TPR (0.81 versus 0.26--0.49), lower TNR (0.70 versus 0.96--0.99), higher TSS (0.51 versus 0.22--0.47), and higher HSS (0.39 versus 0.12--0.22).  Exceptions to this trend are the method of \citet{song2009}, which was applied on a very small dataset of 65 samples, and the method of \citet{ahmed2013} which has a higher HSS due to lack of dataset balancing.  \citet{ahmed2013}, \citet{yuan2010}, and~\citet{welsch2009} do not appear to balance their datasets prior to classification, which likely skews their accuracies in favor of the majority negative class.  Compared to those works wich do use temporal information~\citep{huang2010,yu2010a,yu2010b}, we find our method to have lower TPR (0.81 versus 0.85--0.95) and lower TNR (0.70 versus 0.88--0.98).  There are three potential sources for this difference in performance: the different flare magnitudes, the use of temporal features, and different size datasets. We will consider the implementation of temporal features as well as study the effect of different flare magnitudes in future work as we will discuss in Sect.~\ref{sec:conclusions}.

\section{Conclusion and future work}
\label{sec:conclusions}
We used a large set of LOS magnetograms, including ARs which ultimately flared and control ARs which did not flare. We extracted 38 different features related to the complexity of each AR magnetogram. These features resulted from an analysis of the total flux, spatial gradient, NL, flux evolution, and wavelet decomposition. This is the largest scale study carried out to date, in terms of combining a large number of features and a large dataset. An RVM standard pattern recognition framework was used to classify whether the given AR will produce a solar flare. In general, we achieved TPRs of $\sim$0.8 and TNRs of $\sim$0.7. These rates correspond to a TSS of approximately 0.5. In comparison to other studies of flare prediction from static images ~\citep{ahmed2013,yuan2010,huang2010,yu2010a,yu2010b,welsch2009} (and as summarized in~\citet{bloomfield2012}), we find our classification performance to be higher for TNR and lower for TPR. However, in a comparison to studies that use temporal features, the TPR and TNR discovered here are slightly lower. The TNR and TPR do not vary much when predicting over the 2-24 hour window.

Upon ranking, features related to magnetic gradients are associated with the best predictive ability. Features related to power at various wavelet scale decompositions also feature in the top 5. This agrees with and improves upon previous work, where the size scale of the neutral line was studied \citep{ireland2008}.

The large size of this study, compared to previous work, only resulted in small improvements over previous work. This naturally leads us to question where future advances can be made. Clearly, just adding in more data and more features is not necessarily the best approach. It has always been clear that while the photospheric magnetic field governs the coronal non-potentiality (and hence likelihood to produce a solar flare), photospheric magnetic field information alone is not sufficient to determine coronal structure. Chromospheric, and eventually coronal, magnetic field is required. In addition, we emphasize that this type of study is only measuring a proxy of the magnetic energy {\em build up}. We are still lacking observational details on why energy is released at any particular point in time. It is also unclear, both observationally and theoretically as to how much (i.e., what fraction) of stored energy is released \citep{mcateer2007, mcateer2013c}, and how this is distributed between thermal emission, non-thermal emission, and bulk motions \citep{emslie2012}. With this in mind, we may have discovered the natural limit of the accuracy of flare predictions from these large scale photospheric studies.

However, some further advances can be made with current data. We plan for three further investigations related to the features themselves. First, we will implement fractal dimension features, to include computation of the fractal dimension of the NL and features related to the grayscale fractal dimension of the magnetogram itself.  Fractality and multifractality has been shown to be a highly discriminative feature in other application domains~\citep{spillman2004, mcateer2013} and warrants  further investigation~\citep{conlon2008, mcateer2015, mcateer2015b}.  As a second investigation related to the complexity features, we will consider the use of feature selection methods.  While we have considered arbitrary feature subsets according to the image processing methods (e.g., gradient or NL), automated methods can determine optimal (or close to optimal) feature subsets~\citep{pudil1994}. This analysis will allow insight into the specific physical processes that directly precede flares. As a third exploration, we will consider the implementation of temporal features to characterize the change in appearance of active regions leading up to a flare.

We also plan for four further investigations related to the pattern recognition aspects of our work.  First, we will repeat this experiment for a variety of flare sizes (e.g., C1.0, C5.0, M1.0, M5.0, X1.0) to study and mitigate the bias associated with high solar backgrounds. Second, we will investigate other means of classifying our unbalanced dataset.  This work used a cross-validation framework in which the dataset was subsampled to yield equal contribution from flaring and non-flaring regions.  As the different errors (FP and FN) have very different implications in flare forecasting, we can implement a cost matrix which applies a different penalty to the two different errors.  This may allow us to tune the performance to a better suited level for flare prediction. Third, we will implement these features in a regression analysis (using RVMs) where we will predict \emph{when} a flare will occur rather than the binary decision that a flare will occur within some timeframe.  This will provide further insight into the predictive time windows associated with flare prediction, and which features may be more applicable across the different predictive time windows. Fourth, we will analyze the classification and regression frameworks for prediction of other solar eruptive events often coincident with solar flares, including coronal mass ejections and solar energetic particles.

\begin{acknowledgements}
The authors gratefully acknowledge an NMSU Vice President for Research Interdisciplinary Research Grant, NSF PAARE grant AST-0849986, NASA EPSCoR grant NNX09AP76A, and NNH09CE72C, all of which helped support this work. One of us (JMA) was partially supported by a National Science Foundation Career award NSF AGS-1255024, and NASA contracts NNH12CG10C and NNX13AE03G.
\end{acknowledgements}


\bibliographystyle{aa}
\bibliography{main}
\end{document}